\UseRawInputEncoding
\documentclass[lineno]{jfm}

\usepackage{graphicx}
\usepackage{epstopdf,epsfig}
\usepackage{newtxtext}
\usepackage{newtxmath}
\usepackage{natbib}
\usepackage{hyperref}
\hypersetup{
    colorlinks = true,
    urlcolor   = blue,
    citecolor  = black,
}

\newcommand{\RomanNumeralCaps}[1]
\linenumbers

\title{Insights into the shockwave attenuation in miniature shock tubes}

\author{Janardhanraj S.\aff{1}
	Abhishek K.\aff{1}
	\and Jagadeesh G.\aff{1}
	\corresp{\email{jaggie@iisc.ac.in}}}

\affiliation{\aff{1}Department of Aerospace Engineering, Indian Institute of Science, Bengaluru 560012 INDIA}

\begin{document}
\maketitle

\begin{abstract}
Miniature shock tubes are finding growing importance in a variety of interdisciplinary applications. There is a lack of experimental data to validate the existing shock tube flow models that explain the shockwave attenuation in pressure-driven miniature shock tubes. This paper gives insights into the shock formation and shock propagation phenomena in miniature shock tubes of 2mm, 6mm, and 10mm square cross-sections operated at diaphragm rupture pressure ratios in the range 5-25 and driven section initially at ambient conditions. Pressure measurements and visualization studies are carried out in a new miniature table-top shock tube system using nitrogen and helium as driver gases. The experimental findings are validated using a shock tube model explained in terms of two regions; (1) The shock formation region, dominated by wave interactions due to the diaphragm's finite rupture time. (2) The shock propagation region, where the shockwave attenuation occurs mainly due to wall effects and boundary layer growth. Correlations to predict the variation of shock Mach number in the shock formation region and shock propagation region work well for the present findings and experimental data reported in the literature. Similar flow features are observed in the shock tubes at the same dimensionless time stamps. The formation of the planar shock front scales proportionally with the diameter of the shock tube. The peak Mach number attained by the shockwave is higher as the shock tube diameter increases.
\end{abstract}

\begin{keywords}
Authors should not enter keywords on the manuscript, as these must be chosen by the author during the online submission process and will then be added during the typesetting process (see \href{https://www.cambridge.org/core/journals/journal-of-fluid-mechanics/information/list-of-keywords}{Keyword PDF} for the full list).  Other classifications will be added at the same time.
\end{keywords}

{\bf MSC Codes }  {\it(Optional)} Please enter your MSC Codes here

\section{Introduction}\label{sec:intro}
Shock tubes are devices used to generate shockwaves in a safe and reproducible manner in laboratory confinement. A simple shock tube has a high-pressure chamber (known as the driver section) and a low-pressure chamber (known as the driven section) separated by a diaphragm. The diaphragm rupture leads to a shockwave formation, which propagates down the driven section of the shock tube \citep{Bradley,gaydon}. Shock tubes are commonplace in research laboratories to facilitate chemical kinetic studies, supersonic and hypersonic investigations. Though seemingly simple devices, shock tubes are a subject of intense scrutiny with several unanswered questions. Shock formation and attenuation in shock tubes have been areas of long-standing research. Over the years, there have been reports that address these aspects, albeit at low initial pressures in the driven section and using limited experiments. Emerging transdisciplinary applications of shockwaves have been demonstrated at initial ambient conditions in miniature scales. Shockwave-assisted applications using miniature shock tubes in needle-less drug delivery \citep{Jana2017}, bacterial transformation \citep{datey2017}, and suppression of cavity noise \citep{Rama2010} typifies the growing interest in this area. Miniature shock tubes driven by lasers have been reported for use in hydrodynamic studies and spectroscopy studies \citep{Busq2010,Zvorykin2000}.
 High repetition miniature shock tubes have been developed using high-speed pneumatic valves for high-pressure and high-temperature reacting systems \citep{tranter2013,Lynch2015}. Extensive experiments and exhaustive analysis are essential to provide insights into the shockwave formation, propagation, and attenuation dynamics in miniature shock tubes operated at initial ambient conditions for further development of shockwave-based applications.

A brief review of the studies on the shockwave formation and attenuation is useful. \citet{glass1955} were among the first to report that there are two main reasons for attenuation in a shock tube, namely, the shock formation process and the effects due to the shock tube walls. But there was a lack of experimental data to validate this model for different operating conditions in shock tubes. \citet{white1958} proposed a model to account for the finite time taken by the diaphragm to rupture in large diameter shock tubes operated at high diaphragm pressure ratios. He experimentally showed that the shockwave velocity could be higher than the values predicted by the one-dimensional inviscid shock tube theory. \citet{emrichcurt1953} predicted that stronger shockwaves attenuate faster than weaker shockwaves, and the attenuation per unit length is almost independent of distance. \citet{Mirels1963,Mirels1964} and \citet{emrichwheel1958} helped understand the wall effects in shock tube flow by accounting for the boundary layer development behind the moving shock front.  \citet{ikui1969_2} provided an improved multistage approach to White's model for the shock formation process. They also proposed a relation between shock formation distance($x_f$) and hydraulic diameter ($D$) of the shock tube as $x_f \propto D^{0.88}$ \citep{ikui1969}. \citet{low1974} described the diaphragm opening process for different materials used as diaphragms. They also clearly defined the shock formation distance as the distance from the diaphragm location where the shock speed reaches its maximum value. \citet{low1976} presented a qualitative description of the diaphragm opening process in shock tubes, which is an important parameter that decides the shock formation distance. Their experimental results showed good agreement with the relation present by \citet{drewry1965} for the diaphragm opening time ($t_{op}$) given by,
	\begin{equation}
	t_{op} = K\sqrt{\frac{\rho.b.th}{P_4}}
	\label{e1_op}
	\end{equation}
	where $\rho$ is the density of the material, $b$ is the length of the petal base, $th$ is the petal thickness, $P_4$ is the bursting pressure, and $K$ is taken as 0.93. \citet{simp1967} represented the shock formation distance as, $x_f=K_1.V_{Smax}.t_{op}$ , where $V_{Smax}$ is the maximum shock velocity, $t_{op}$ is the opening time of the diaphragm, and $K_1$ is the constant of proportionality. \citet{low1976} reported that the value of $K_1$ mostly lies between 1 and 3, and sometimes greater than 3. \citet{ikui1979} observed that the shockwave becomes planar at a distance, which is about a fifth of the shock formation distance. \par

Studies in a microchannel with a hydraulic diameter of 34$\mu m$ revealed that shockwave propagation at micro-scales exhibits a behavior similar to that observed in larger-scale facilities operated at low initial pressures \citep{Mir2012}. \citet{Mir2009} performed experiments in a 5.3mm diameter shock tube at low initial driven section pressures (typically $P_1 < 100mbar$) and showed good qualitative agreement with a flow model based on a scaling parameter $Scl=Re'D/4L$, where $Re'$ is the characteristic Reynolds number based on the driven gas, $D$ is the hydraulic diameter of the tube and $L$ is the characteristic length. \citet{Arun2012,Arun2012_3,Arun2013,Arun2013_1,Arun2014} presented a series of numerical reports on the shock formation process in micro-shock tubes due to the gradual rupture of the diaphragm when operated at very low initial pressures in the driven section. They reported that the shockwave attenuation is significantly higher in micro-shock tubes as compared to macro-scale shock tubes. \citet{Sun2001} reported that viscous effects in channels the height of which is below $4mm$ become noticeable even at atmospheric pressure. \citet{Park2012} reported that at the same initial conditions, the shockwave attenuation in a $3mm$ shock tube is more significant than in a $6mm$ shock tube. \citet{Zeitoun2006} used 2-D Navier-Stokes computations with slip velocity and temperature jump boundary conditions to predict flow in micro-scales. Numerical studies by \citet{Ngomo2010} show that the flow in microchannels shows a transition from an adiabatic regime to an isothermal regime. A one-dimensional numerical model to predict micro-scale shock tube flow was presented by \citet{Mir2009} by integrating the three-dimensional diffusion effects as sources of mass, momentum, and energy in the axial conservation equations. \citet{giordano2010} studied the transmission of weak shock waves through $1.02mm$ and $0.48mm$ miniature channels. \citet{Mir2013} later complemented their experimental results with a Navier-Stokes model, which assumes a no-slip isothermal wall boundary condition. Recently, pressure measurements and particle tracking velocimetry were made in a $1mm$ square shock tube operating at diaphragm pressure ratios of 5 and 10 \citep{Zhang2016}. \par

The present study highlights the shock attenuation phenomena in miniature shock tubes for operating conditions similar to the practical scenario used in the shockwave-based applications. The most commonly used diameters of shock tubes for transdisciplinary applications are the range of 10mm or lower. Also, the driven section of the shock tube is generally kept at ambient conditions. Therefore, experiments are performed in a 2mm, 6mm, and 10mm shock tube with a square cross-section in a unique table-top shock tube facility in the present work. A square cross-section is chosen to facilitate visualization studies of the driven section of the shock tube. For the first time, the shock tube's entire driven section is visualized to study the shockwave formation and attenuation. Nitrogen and helium are used as driver gases, and the diaphragm pressure ratio is also varied. The obtained results from the experiments are compared with analytical and computational models reported in the literature. The wave phenomena that occur immediately after the diaphragm rupture in the driven section of the shock tube are discussed in detail. The formation of Mach stem and triple point due to the diaphragm's finite opening time and the evolution of a planar shock front is presented. Correlations are developed based on the experimental and numerical findings to predict the shock Mach number's variation along the entire length of the driven section. These correlations perform satisfactorily for the present experimental findings and data reported in the open literature.

\section{Experimental Methodology}\label{sec:methodology}

\subsection{Shock tube facility}

\begin{figure}
  \centerline{\includegraphics[scale=0.55]{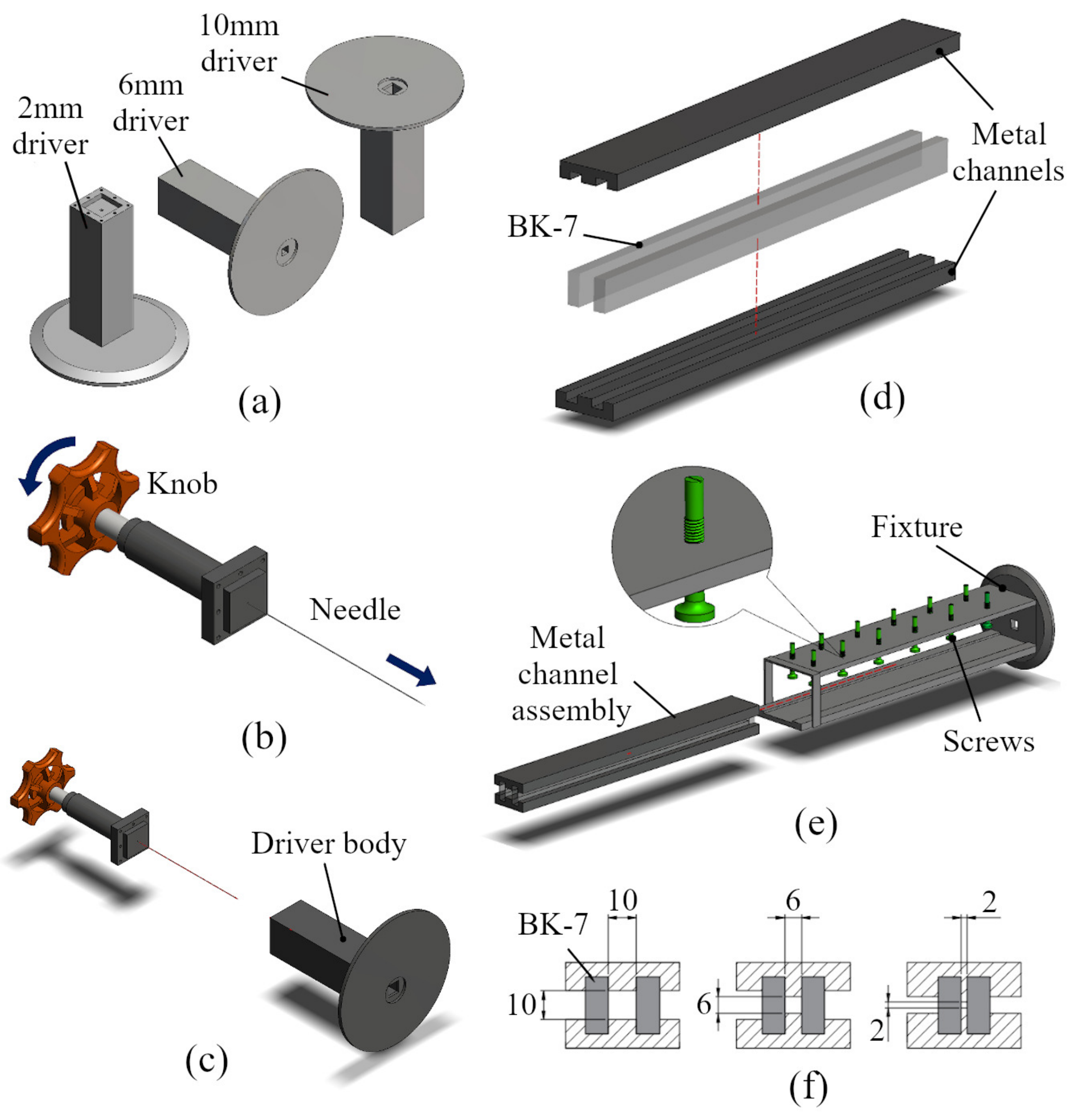}}
  \caption{(a) CAD models showing the driver sections of the three shock tubes. (b) Diaphragm puncture mechanism used in the driver section. (c) Assembly of the diaphragm puncture mechanism in the driver section. (d) Exploded view showing the assembly of the BK-7 glass slabs. (e) Assembly of the driven section of the shock tube. (f) Cross-section showing the driven section of the three shock tubes.}
\label{fig1}
\end{figure}

	The present study uses a unique table-top shock tube designed and built in-house at the Laboratory for Hypersonic and Shockwave Research (LHSR), Indian Institute of Science (IISc), Bengaluru, India. This particular experimental setup, which has a quick-changing diaphragm mechanism, is used for pressure measurements and visualization studies in miniature shock tubes. The experimental setup is similar to the one described in a previous study \citep{jana2016} with additional features. The shock tube's internal dimension can be changed to either a 2mm or 6mm or 10mm square cross-section. The length of the driver section is 100mm, while the driven section has a length of 339mm. Three separate driver sections of 2mm, 6mm, and 10mm square internal cross-section of length 100mm are fabricated (see Figure \ref{fig1}a).
There is also a diaphragm puncture mechanism incorporated in the shock tube (see Figure \ref{fig1}b). A long needle is connected to the driver section's rear end to puncture the diaphragm (see Figure \ref{fig1}c). A knob moves the needle in the forward direction (a total distance of about 5mm) when rotated. This mechanism can be used interchangeably in the three different driver sections. Two optical quality BK-7 glass slabs form the walls of the driven section of the shock tube (see Figure \ref{fig1}d). Metal channels restrict the motion of the BK-7 glass in the lateral direction. The driven section arrangement encloses either a 2mm by 2mm or 6mm by 6mm or 10mm by 10mm cross-section (see Figure \ref{fig1}f). A separate fixture houses the entire assembly of the driven section. Adjusting screws on top of the fixture holds the driven section in place firmly (see Figure \ref{fig1}e). The fixture's metal frame obstructs a small region covering a distance of 4mm immediately after the diaphragm location and another region covering a distance of 35mm towards the end of the driven section. The total unobstructed length of the driven section that can be viewed from a direction perpendicular to the shock tube axis is 300mm. During pressure measurements, the BK-7 glasses are replaced by a composite plate that houses the pressure sensors, the details of which are given in the subsequent section. All components are made of stainless steel (SS-304 grade). The tolerance of the machined components is $\pm0.02mm$.

\subsection{Measurement of pressure}

\begin{figure}
  \centerline{\includegraphics[scale=0.09]{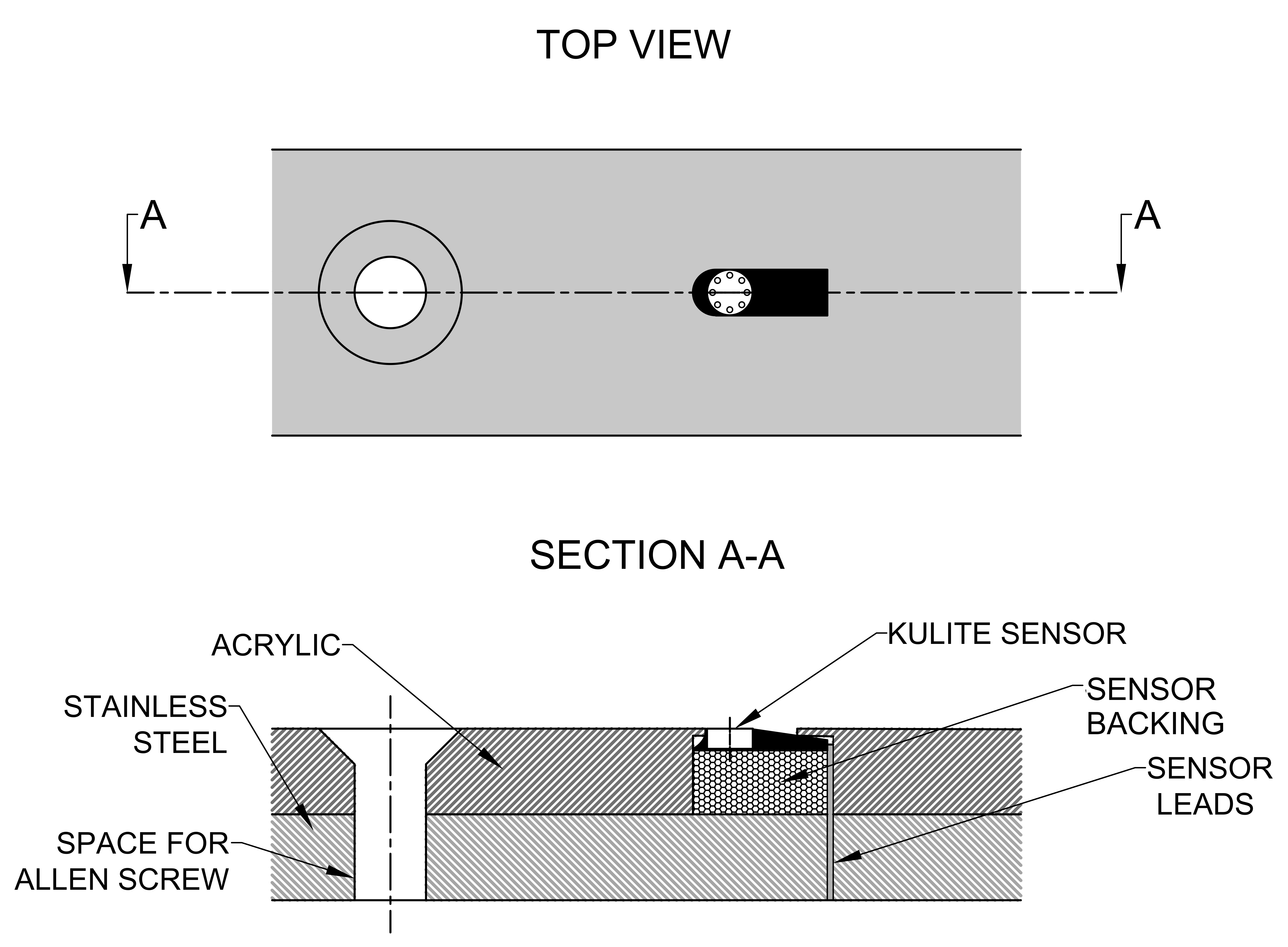}}
  \caption{A 2-D drawing showing the top view and cross-sectional view of the composite plate that accommodates the ultra-thin miniature pressure transducers.}
\label{fig2}
\end{figure}

The pressure of the shock tube's driver gas is measured using a digital pressure gauge (SW2000 series, Barksdale Control Products, Germany) with an operating range of 0-50.0 bar and a least count of 0.1 bar. The pressure histories inside the driven section of the shock tube are obtained using the ultra-miniature pressure transducers (LQ-062 series, Kulite Semiconductor Products Inc., USA). The sensor's natural frequency is 300 kHz, and the rise time is typically 1.3 $\mu$s. The sensing region of the transducer has a diameter of 1.6mm. These sensors are mounted on a composite plate arrangement that ensures that the sensing surface is flush with the shock tube walls. The sensors were installed according to the manufacturer's instructions. Figure \ref{fig2} shows the top view and the cross-sectional view of the composite plate. An acrylic plate and a stainless-steel plate are sandwiched together.
The acrylic plate is used near the sensor to avoid any electrically conducting material. The inside of the acrylic plate is modified to expose only the sensing surface of the pressure sensor. The leads of the sensor are drawn out through a small hole made in the stainless-steel plate. Suitable backing is given to the sensor to ensure the sensing surface remains flush to the plate surface. A screw fastens the acrylic and the stainless-steel plate together. Two miniature pressure sensors are mounted in the shock tube's driven section at a distance of 291mm (sensor 1) and 331mm (sensor 2) from the diaphragm location. A signal conditioning rack (DEWETRON GmbH, Germany) is used for data acquisition from the miniature transducers. The signals are obtained without any amplification from the signal conditioner. A 300 kHz filter is used to cut off the unwanted frequencies embedded in the signal.

\subsection{Shadowgraphy}
	The shockwave propagation in the driven section of the shock tube is captured using a high-speed shadowgraphy technique \citep{gary2001}. The setup comprises of a high-speed camera (Phantom V310, Vision Research, USA) with a maximum acquisition rate of 500,000 frames per second at lowest resolution and a minimum exposure time of 1$\mu$s, concave mirrors of diameter 300mm and focal length 3m, a 5-Watt single LED light source and a digital signal generator (Stanford Research, USA). Since the spanwise length of the shock tube (i.e., 2mm, 6mm, or 10mm) is small compared to the axial length of the driven tube (300mm in length), the aspect ratio (the proportion between the width and the height) of the observation window is very high. Hence, the driven section is divided into two parts, and the portions are visualized separately. Two thin threads are tied around the driven section assembly at distances of 140mm and 160mm from the diaphragm location. PART 1 is the region from the diaphragm location to a distance of 160mm along the driven section. PART 2 is the region from the 140mm mark to a distance of 300mm along the driven section.
	
\subsection{Operating conditions}
	The driven section of the shock tube for all the experiments is left open to the atmosphere. For all the calculations, the $P_1$, $T_1$ and $\rho_1$ is taken as 1 bar, 298K and 1.2 kg/m$^3$ respectively. $P_1$, $T_1$, and $\rho_1$ indicate the initial pressure, temperature, and density of the driven gas, respectively. Cellophane of 40$\mu$m thickness is used as the diaphragm for all experiments. Nitrogen and helium are used as driver gas. The diaphragm rupture pressure ratio ($P_{41}$ = $P_{4}/P_{1}$) is varied between 4.9 and 26.2 for nitrogen as driver gas while it is between 5.1 and 25.8 for helium as driver gas. The range of diaphragm rupture pressures was chosen to avoid potential side effects like a very slow opening or improper opening of the diaphragm. The characteristic Reynolds number for the present conditions is in the range of 45,557 - 227,783. The characteristic Reynolds number is given by, $Re'=(\rho_1 a_1 D)/\mu_1$,  where $a_1=346$m/s and $\mu_1=1.8\times10^{-5}$kg/(m.s) is the speed of sound and dynamic viscosity of driven gas respectively. While performing the visualization experiments, the driver gas is nitrogen, driven gas is air at atmospheric conditions, and $P_{41}$ is 15.
	
\subsection{Uncertainty}
	During experiments, care is taken to ensure that the measurements are made carefully to ensure a high degree of confidence. The uncertainty in the measured quantities is calculated from the different sources of errors and their propagation \citep{taylor1982}. The uncertainty in the measured values of pressure using the ultra-thin miniature pressure transducers is $\pm4\%$. The rise time of the complete system used to acquire the pressure signals is 1.5$\mu$s, which is much lower than the rise time of the shockwave. The shockwave rise time is evaluated by measuring the time elapsed as the initial ambient pressure rises to the peak pressure represented by P$_{21}$. In the present experiments, the shockwave's rise time is about 50$\mu$s, as seen in the plots shown in Figure \ref{fig4}. The uncertainty in the measured shock speed using the time-of-flight method from the signals obtained using the ultra-miniature pressure transducers is $\pm 1\%$. The uncertainty in tracking the location of the shockwave from the high-speed shadowgraphs is $\pm 0.2\%$.

	\section{Experimental results}\label{sec:exp_results}
	\subsection{Pressure measurements}

 \begin{figure}
  \centerline{\includegraphics[scale=1.75]{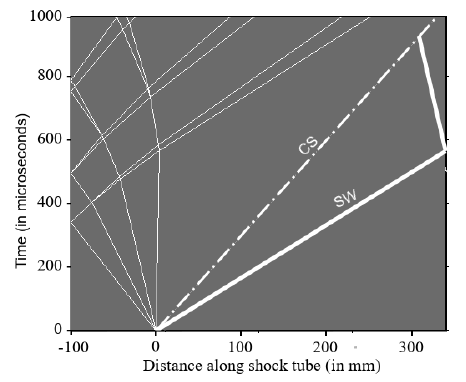}}
  \caption{A wave diagram for the shock tube conditions used in the experiments showing the shockwave (SW), contact surface (CS) and rarefaction waves (thin lines).}
\label{fig3}
\end{figure}

\begin{figure}
  \centerline{\includegraphics[scale=0.6]{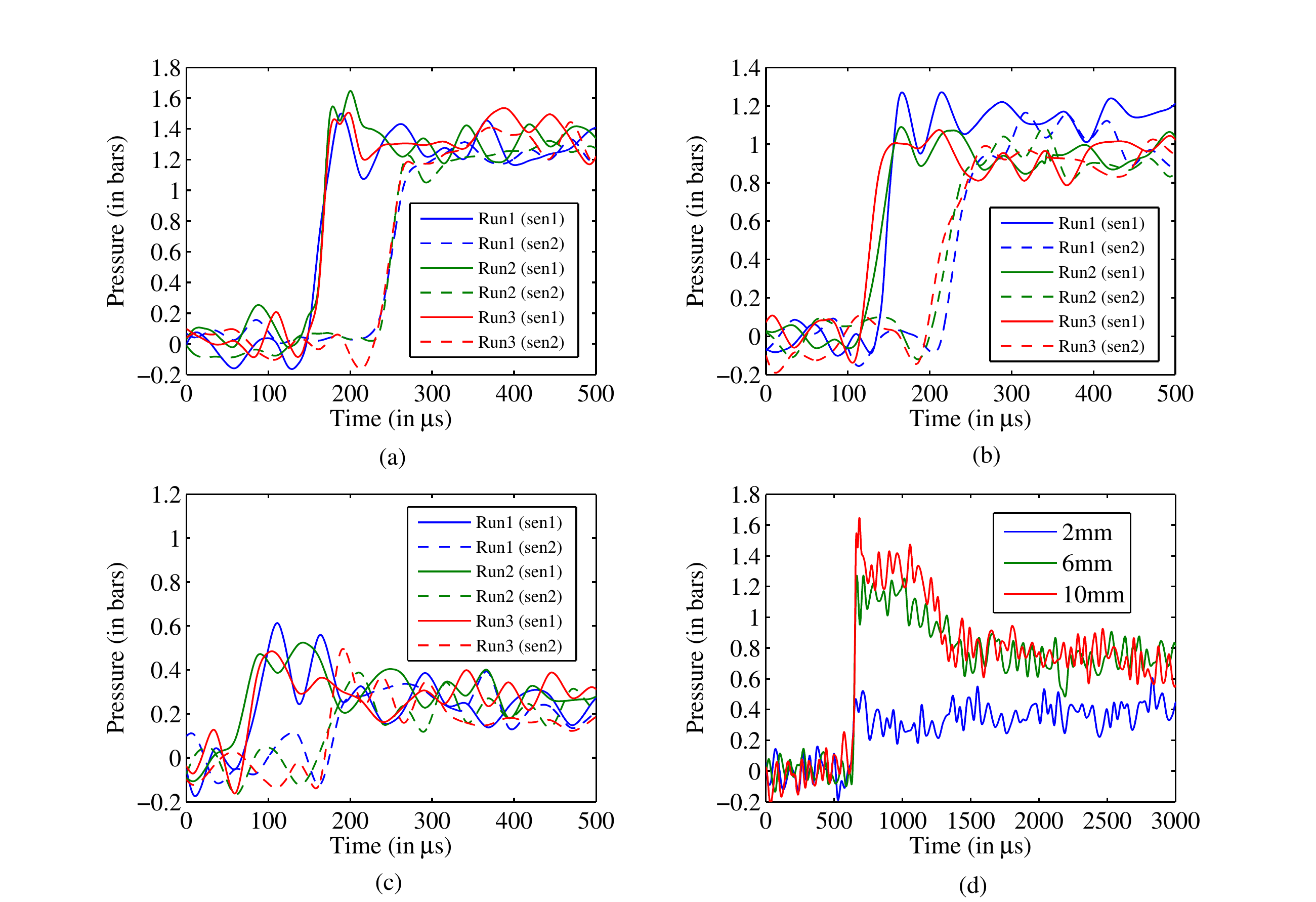}}
  \caption{Plots showing the repeatability of the pressure signals for $P_{41}=15$ in the (a) 10mm shock tube, (b) 6mm shock tube, and (c) 2mm shock tube. (d) A plot showing the comparison of the pressure signals in the three shock tubes.}
\label{fig4}
\end{figure}

The repeatability of the pressure signals is ensured by performing experiments at $P_{41}=15$ in all the three shock tubes. Figure \ref{fig3} shows the wave diagram using 1-D inviscid shock tube theory for $P_{41} = 15$ and driver gas as nitrogen. The 1-D inviscid shock tube theory is routinely used for calculating the various output shockwave parameters in a shock tube. This theory does not account for the shock tube's shape and the diameter. It assumes inviscid-adiabatic flow with ideal gas behavior, instantaneous diaphragm rupture, and thermal equilibrium. It also assumes the shockwave formation at the diaphragm station, hence neglecting the diaphragm opening time and the shock formation process. It also neglects the surface roughness of the shock tube walls and mass diffusion to these walls. However, the 1-D inviscid shock tube theory forms a good reference for the calibration of shock tubes. The figure \ref{fig3} shows that the reflected expansion waves from the driver end wall do not interact with the contact surface or the shock front. Therefore, there is no disturbance in the flow behind the shock front due to the wave interactions. Figure \ref{fig4}a, \ref{fig4}b and \ref{fig4}c show the plots of the pressure signals for $P_{41} = 15$ obtained in three separate runs for the 10mm, 6mm and 2mm square cross-section shock tubes respectively. Nitrogen is used as the driver gas in all these experiments. The shock Mach number and pressure behind the shock front obtained from the 1-D inviscid shock tube theory are 1.73 and 3.34 $bar$, respectively. The pressure signals at the two different sensor locations are also plotted in the figures. The ultra-miniature pressure sensors give the gauge pressure as the output. Therefore, the atmospheric pressure value has to be added to these pressure values. The figure \ref{fig4}d shows the comparison between the pressure behind the shock front for the three shock tubes. After the initial shock front, the constant pressure region is followed by a drop in pressure (after about 500 microseconds from the initial shock front) in the 6mm and 10mm shock tubes. A similar pressure drop is not observed in the case of the 2mm shock tube. The aspect ratio (ratio of length and diameter) of the driver section is a reason for this observation as the 2mm driver (aspect ratio is 50) is very well supported compared to the 6mm driver (aspect ratio is 16.67) and 10mm driver (aspect ratio is 10). Since the present study focuses on the shock formation and propagation in the shock tube's driven section, the pressure drop (occurring after 500$\mu$s) does not affect the subsequent sections' results.\par

	\begin{table}
		\begin{center}
			\def~{\hphantom{0}}
			\begin{tabular}{lccccccc}
				Configuration & & \multicolumn{3}{c}{Experimental measurements} & & \multicolumn{2}{c}{R-H relations} \\
                  &  \hspace{10pt}  & $P_{21}(sensor1)$   &   $P_{21}(sensor2)$ & $M_S(e)$ &   \hspace{10pt}  & $M_{S1}$ & $M_{S2}$ \\
				10mm shock tube  & & 2.44$\pm$0.05 & 2.17$\pm$0.03 & 1.50$\pm$0.01 & & 1.49$\pm$0.05 & 1.42$\pm$0.05\\
				6mm shock tube  & & 2.04$\pm$0.09 & 1.97$\pm$0.04 & 1.43$\pm$0.04& &1.38$\pm$0.09 & 1.35$\pm$0.04\\
				2mm shock tube & & 1.48$\pm$0.05 & 1.36$\pm$0.07 & 1.13$\pm$0.03& &1.19$\pm$0.05 & 1.14$\pm$0.07\\
			\end{tabular}
			\caption{Values of $P_{21}$ and $M_S(e)$ measured from the pressure signals shown in figure \ref{fig4}. $M_{S1}$ and $M_{S2}$ are the shock Mach numbers calculated from the normal shock relations using $P_{21}$ at sensor 1 and sensor 2 location respectively.}
			\label{t1_rep}
		\end{center}
	\end{table}

	The pressure behind the shockwave ($P_{21}$) is estimated with the condition that there is a maximum of $\pm 5\%$ variation about obtained value. The obtained values of shock Mach number and $P_{21}$ are tabulated in terms of the standard error in Table \ref{t1_rep}. The shock Mach number, $M_S(e)$, is calculated by the time-of-flight method by dividing the distance between the two pressure transducers by the difference in arrival time of the shock front. The variation in the obtained values is represented as the standard error of the mean (SEM). In general, the Rankine-Hugoniot jump relation gives the relationship between $M_S$ and $P_{21}$ as follows,

\begin{equation}
	P_{21} = 1+\frac{2\gamma}{\gamma+1}({M_S}^2-1)
	\label{e2_hug}
\end{equation}

where $\gamma$ is the specific heat ratio of the gas. Using the equation \ref{e2_hug}, the shock Mach numbers, $M_{S1}$ and $M_{S2}$, are estimated from the $P_{21}$ at sensor 1 and sensor 2 locations respectively. $M_S(e)$ and $P_{21}$ are lower than the values obtained using 1-D inviscid shock tube theory. The attenuation in the shock Mach number and pressure behind the shock front is more as the shock tube's internal dimension decreases. Also, $M_{S1}$ and $M_{S2}$ are lower than $M_{S}(e)$ in table \ref{t1_rep} except in the 2mm shock tube case. The shock tube conditions are repeatable in all the cases (the maximum standard deviation is 0.04 for the 6mm square shock tube case). Further, experiments are carried out over a range of diaphragm rupture pressures and for different driver gases. Table \ref{t2_n2} and \ref{t3_he} show the experimental results for different initial pressure in the driver for the 2mm, 6mm, and 10mm square cross-section shock tubes for driver gas nitrogen and helium, respectively. These experiments show the influence of the driver gas. There is a significant drop in the pressure behind the shock front as it travels from sensor 1 location to sensor 2 location. These results are compared to the one-dimensional inviscid shock tube theory and the model proposed by \citet{Broui2003} in section \ref{sec:shock_prop}.

	\subsection{Visualization studies}
	
\begin{figure}
  \centerline{\includegraphics[scale=0.8]{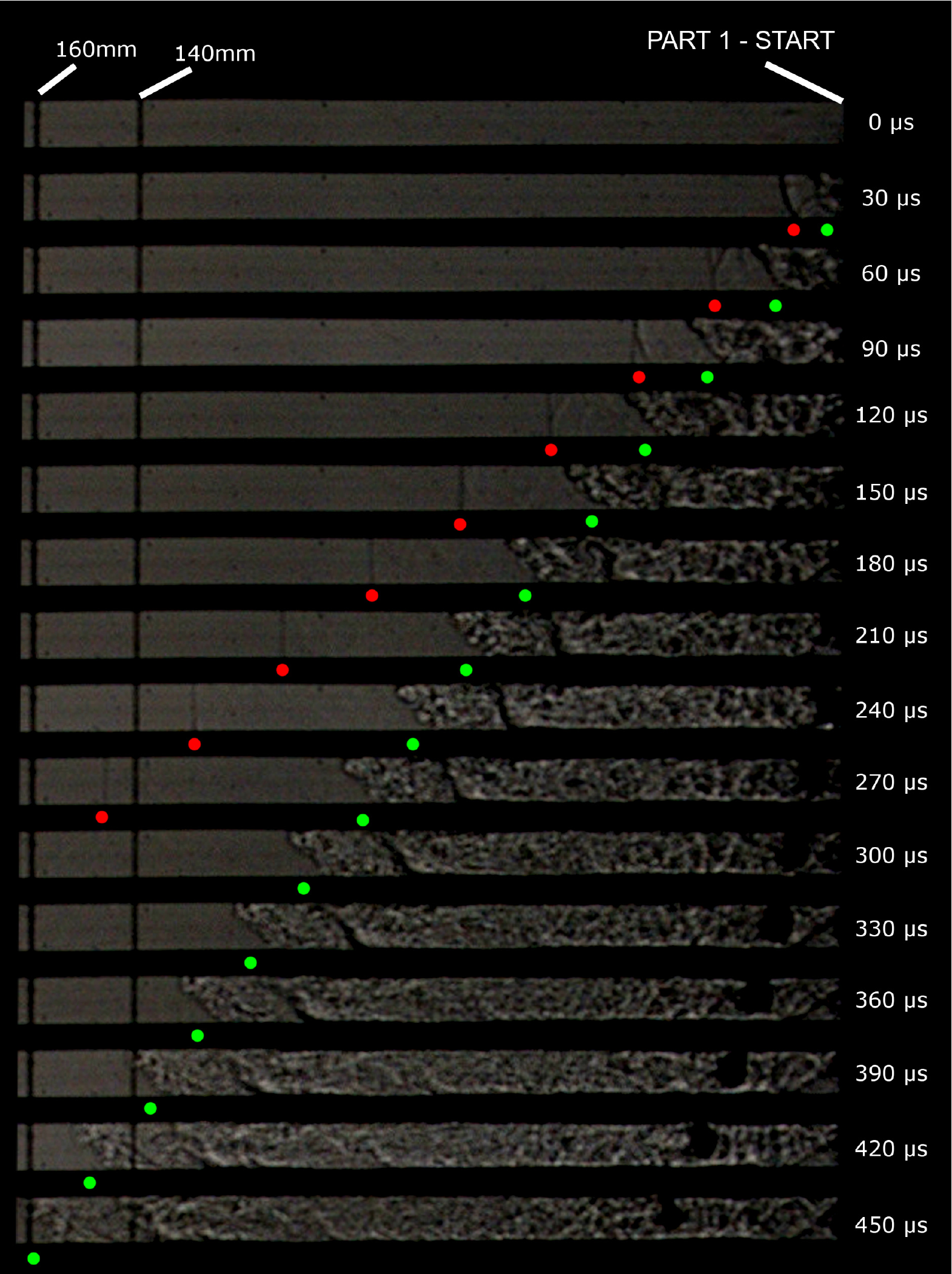}}
  \caption{Sequential shadowgraphs captured of the driven section (part 1) of the 10mm shock tube. Shockwave location indicated by the red dot. Contact surface location indicated by green dot. $P_{41}=15$ and driver gas is nitrogen.}
\label{fig5}
\end{figure}

\begin{figure}
  \centerline{\includegraphics[scale=0.8]{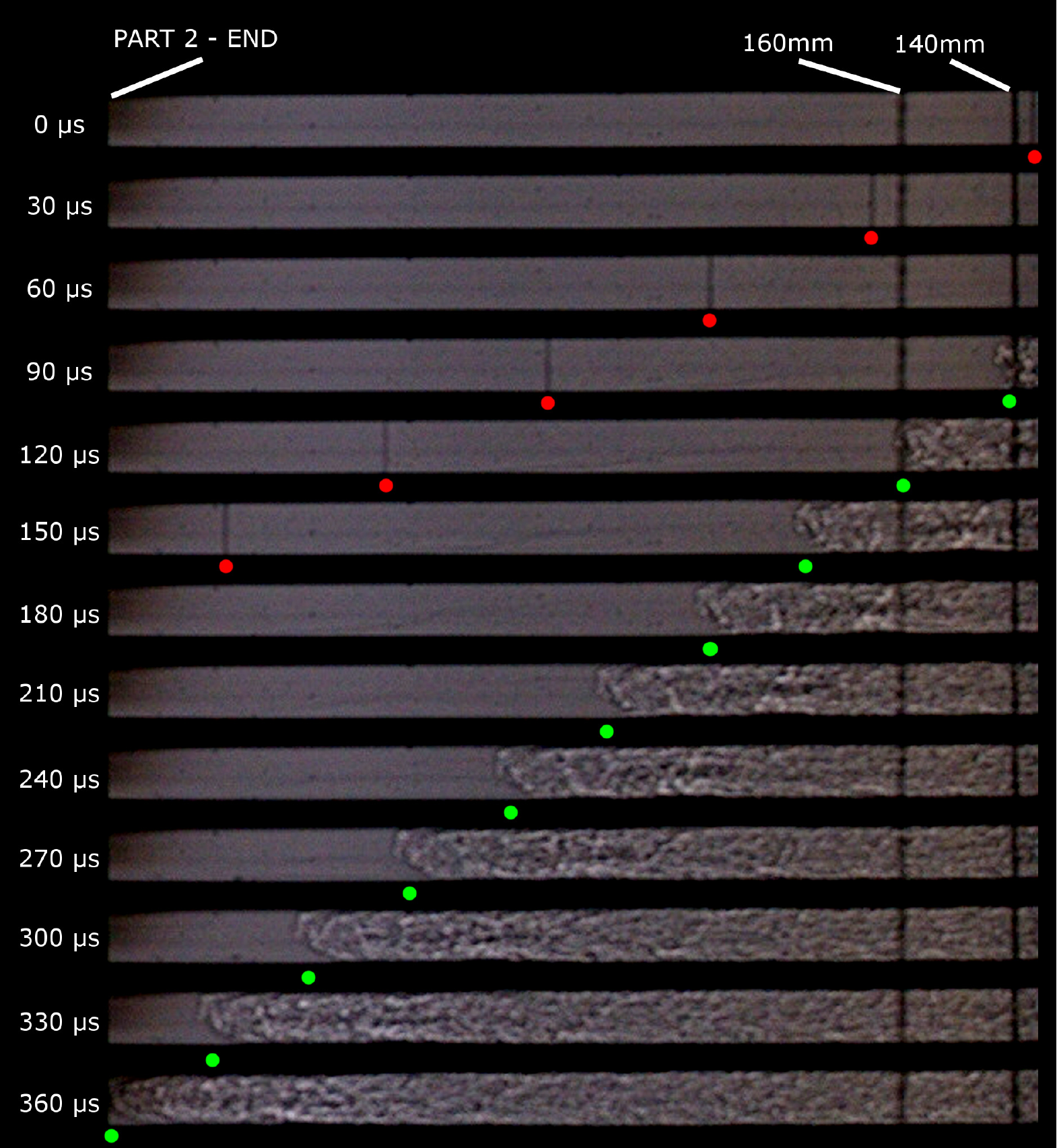}}
  \caption{Sequential shadowgraphs captured of the driven section (part 2) of the 10mm shock tube. Shockwave location indicated by the red dot. Contact surface location indicated by green dot. $P_{41}=15$ and driver gas is nitrogen.}
\label{fig6}
\end{figure}

\begin{figure}
  \centerline{\includegraphics[scale=0.8]{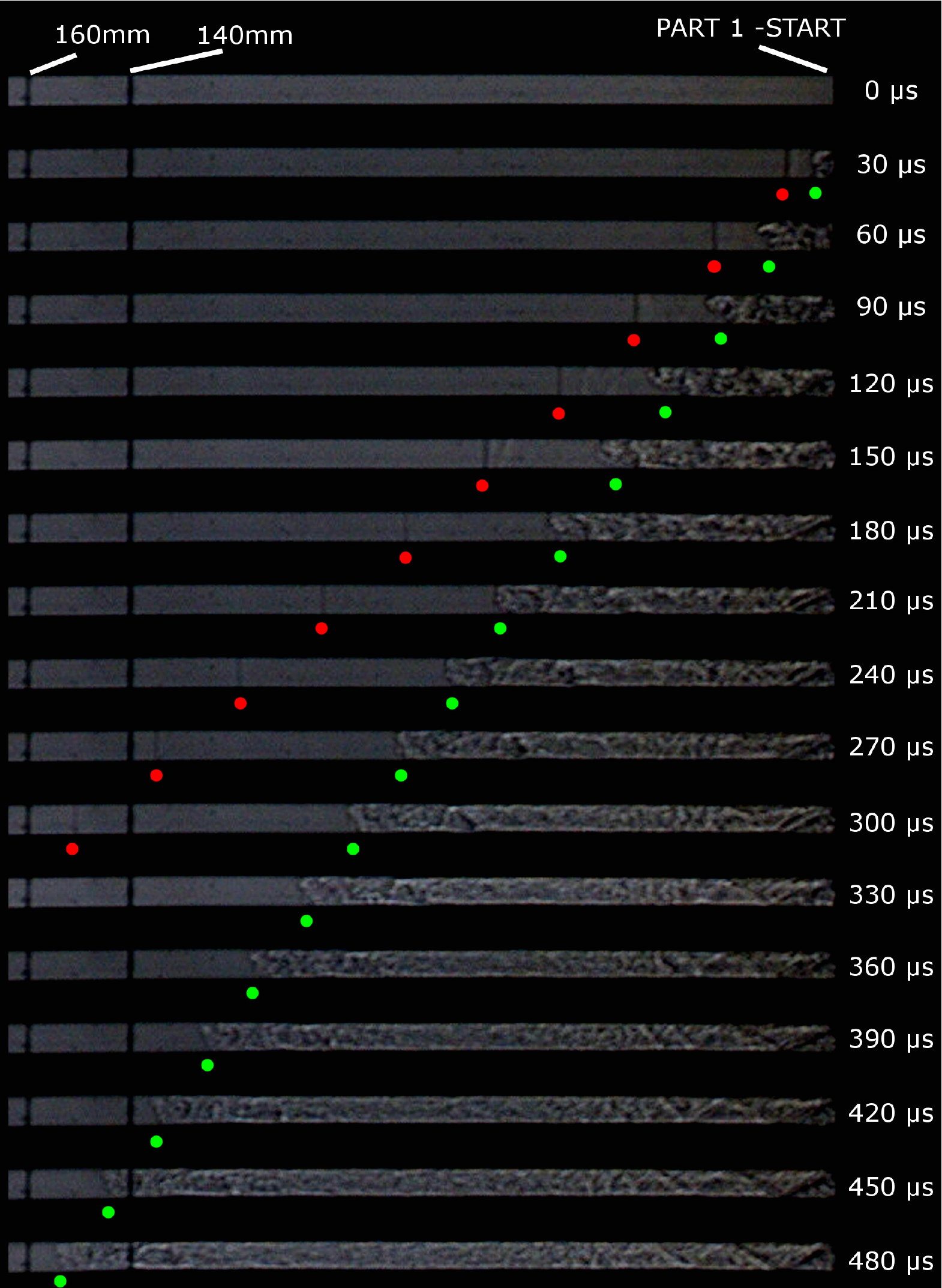}}
  \caption{Sequential shadowgraphs captured of the driven section (part 1) of the 6mm shock tube. Shockwave location indicated by the red dot. Contact surface location indicated by green dot. $P_{41}=15$ and driver gas is nitrogen}
\label{fig7}
\end{figure}

\begin{figure}
  \centerline{\includegraphics[scale=0.8]{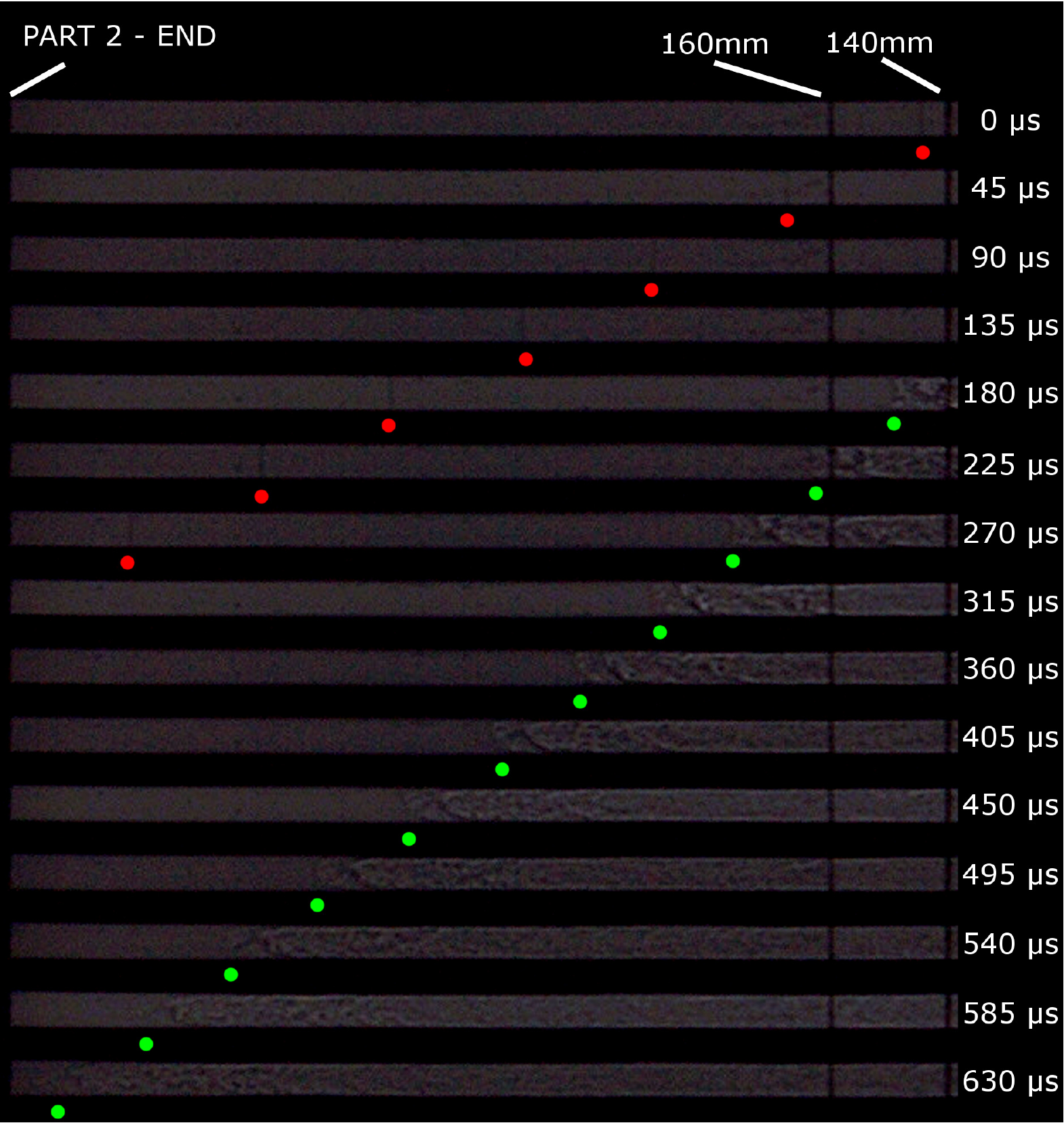}}
  \caption{Sequential shadowgraphs captured of the driven section (part 2) of the 6mm shock tube. Shockwave location indicated by the red dot. Contact surface location indicated by green dot. $P_{41}=15$ and driver gas is nitrogen}
\label{fig8}
\end{figure}

\begin{figure}
  \centerline{\includegraphics[scale=0.8]{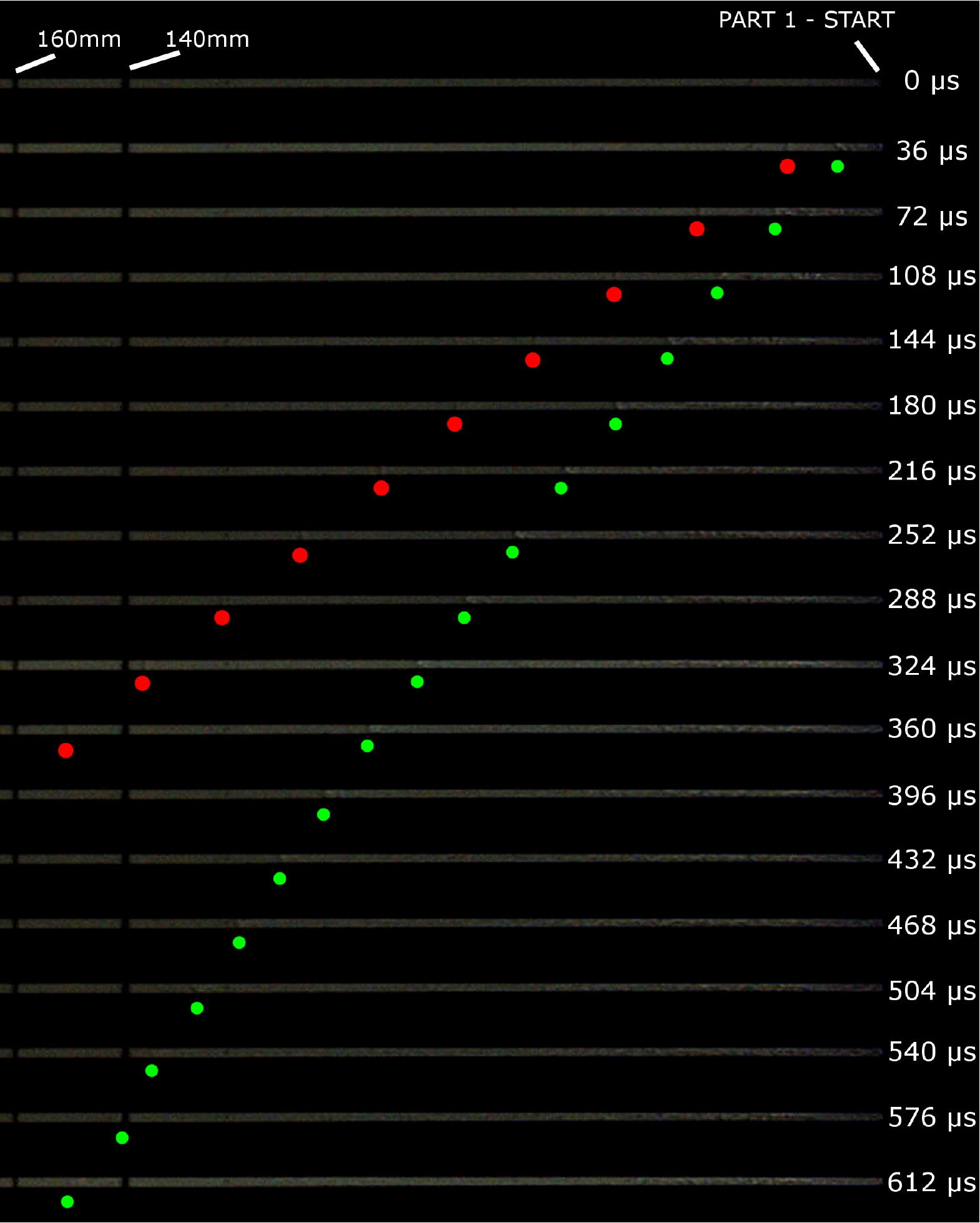}}
  \caption{Sequential shadowgraphs captured of the driven section (part 1) of the 2mm shock tube. Shockwave location indicated by the red dot. Contact surface location indicated by green dot. $P_{41}=15$ and driver gas is nitrogen}
\label{fig9}
\end{figure}

\begin{figure}
  \centerline{\includegraphics[scale=0.8]{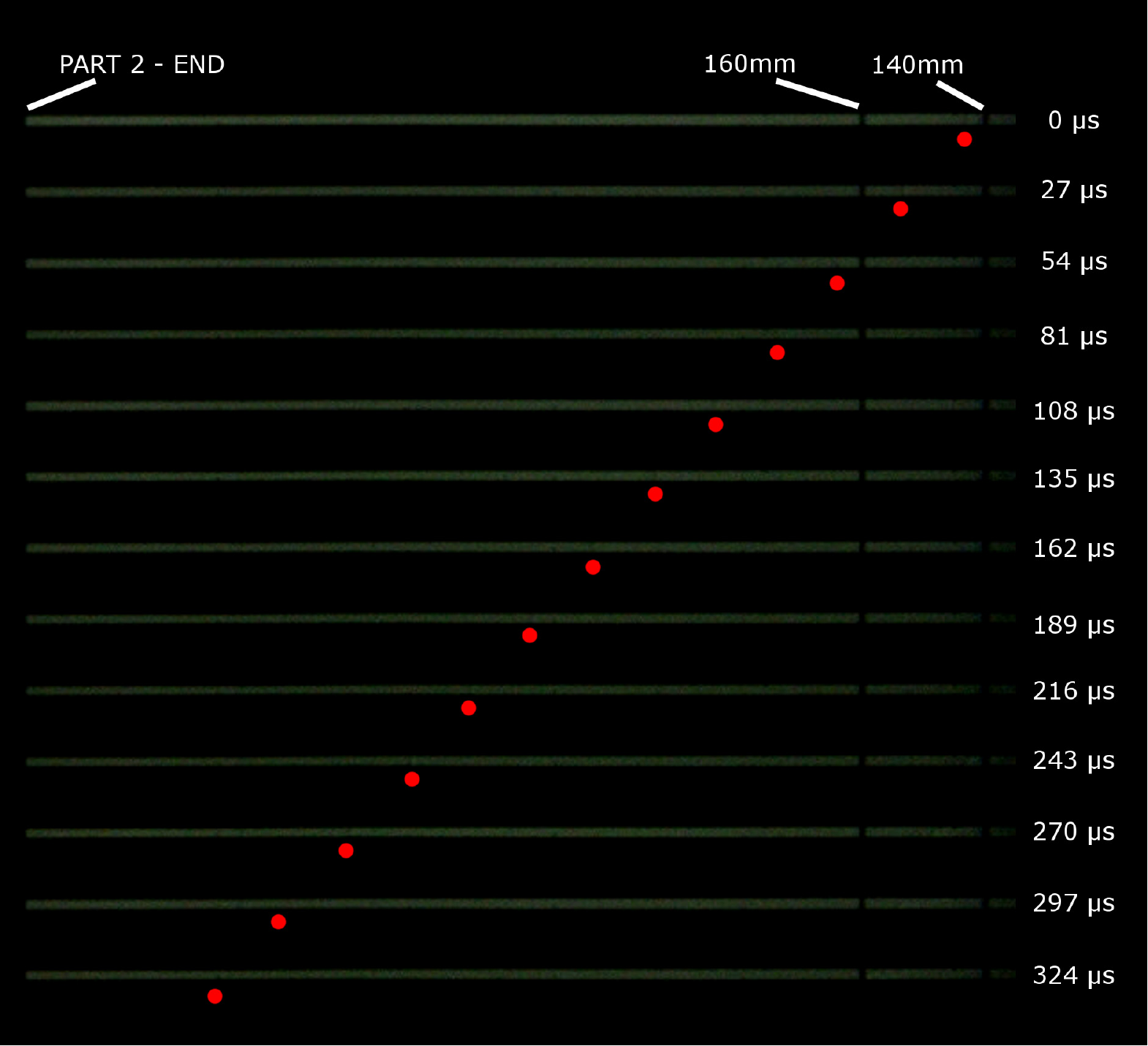}}
  \caption{Sequential shadowgraphs captured of the driven section (part 2) of the 2mm shock tube. Shockwave location indicated by the red dot. Contact surface location indicated by green dot. $P_{41}=15$ and driver gas is nitrogen}
\label{fig10}
\end{figure}

Figure \ref{fig5} shows the sequential images of the 10mm shock tube's Part 1 portion of the driven section for $P_{41}=15$ and nitrogen as driver gas. The images are captured with a frame rate of 97,074 fps at a screen resolution of 560 by 48 pixels. The trigger to the camera is the signal obtained from the pressure sensor at the shock tube's end. The position of the shockwave and the contact surface is indicated for each of the shadowgraphs. Immediately after the diaphragm burst, the curved shock front becomes planar as it travels down the shock tube's driven section. There are also oblique structures visible behind the shock front during the initial frames, which later catch up with the shock front. These compression waves undergo multiple reflections from the walls of the shock tube. The contact surface behind the moving shock front has turbulent structures visible in the images. These turbulent structures can be attributed to the driver gas flow past a partially open diaphragm that leads to a formation of a complex wave system. Figure \ref{fig6} shows the sequential images of PART2 of the 10mm shock tube for the same experimental condition. The images are captured with the same frame rate of 97,074 fps at a screen resolution of 560 by 48 pixels. The timestamps shown in figure \ref{fig5} are independent of those in figure \ref{fig6}. The first image in which the shockwave is seen is given the time stamp $t=0 \mu s$. The figure \ref{fig7} shows the sequential images of Part1 of the driven section of the 6mm shock tube. These images are also captured at a frame rate of 97,074 fps at a screen resolution of 560 by 48 pixels. The shock front is almost planar, while a turbulent contact surface region similar to the 10mm shock tube case is seen. Similar to the 10mm shock tube's shadowgraphs, there are oblique wave structures that later catch up with the shock front. The propagation of the shockwave in Part2 of the driven section of the 6mm shock tube is shown in figure \ref{fig8}. The images are captured at a frame rate of 110,236 fps and resolution of 560 by 40 pixels. Figures \ref{fig9} and \ref{fig10} show the sequential images of Part1 and Part2 of the driven section of the 2mm shock tube respectively. Unlike the 10mm and 6mm shock tube cases, there are many limitations in acquiring good quality shadowgraphs for the 2mm shock tube. The aspect ratio of the driven section of the 2mm square shock tube is very high, and therefore, in the present camera configuration, there are very few pixels that cover the shock tube in the lateral direction. Moreover, the shockwave cannot be tracked towards the end of the shock tube as the wave structures are not distinct. The integration thickness for the shadowgraph is reduced, decreasing the signal-to-noise ratio of the images. The 2mm shock tube images are captured at 110,236 fps and a resolution of 560 by 40 pixels.

\begin{figure}
  \centerline{\includegraphics[scale=0.6]{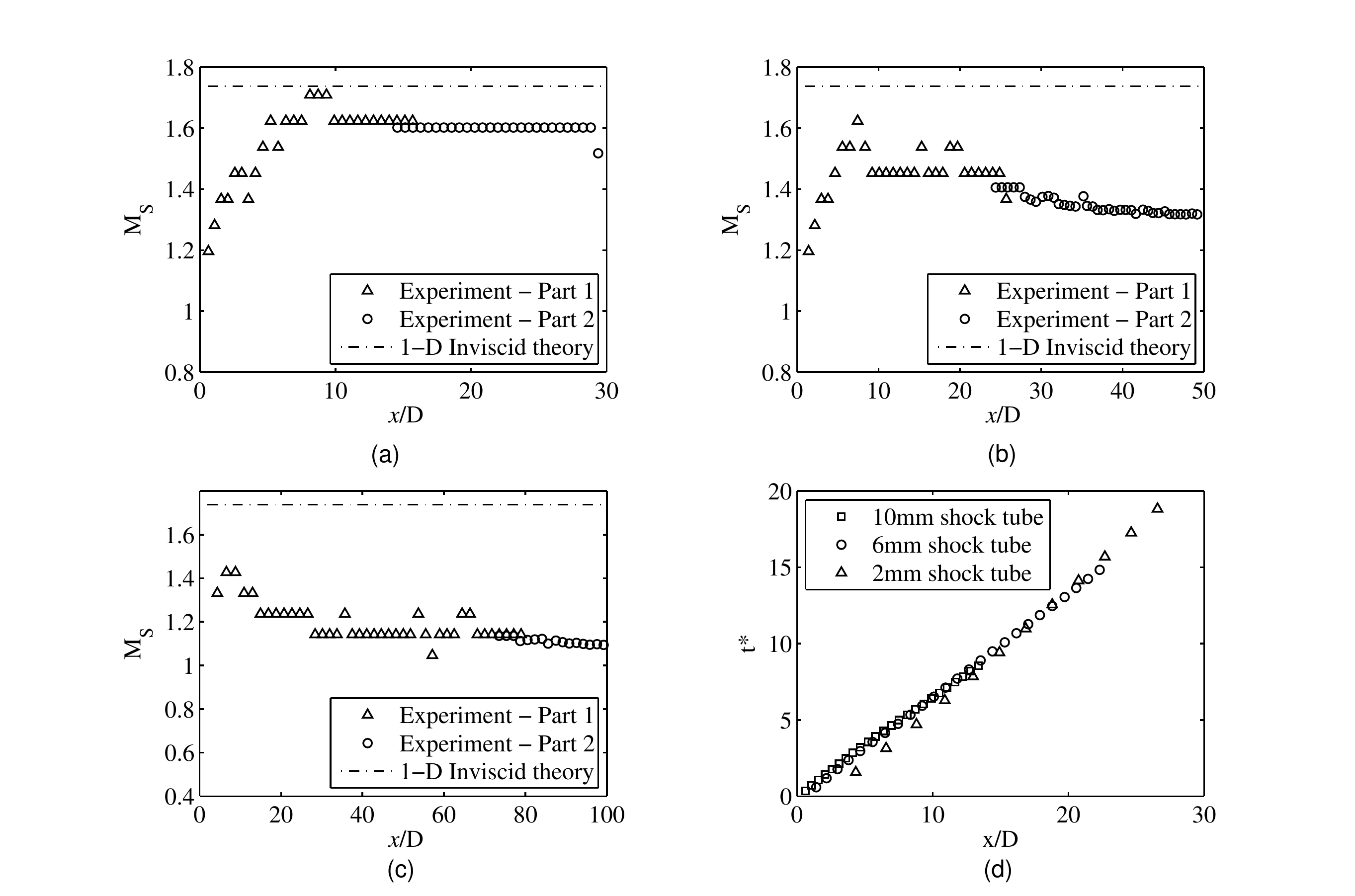}}
  \caption{Plots showing the variation of the shock Mach number along the driven section of the (a) 10mm shock tube, (b) 6mm shock tube, and (c) 2mm shock tube. (d) A plot showing the variation of $t^*$ and $x/D$ in the 10mm, 6mm and, 2mm shock tubes. ($P_{41}=15$ and Nitrogen is driver gas).}
\label{fig11}
\end{figure}

Using the images acquired by the shadowgraph technique in the figures \ref{fig5}, \ref{fig6}, \ref{fig7}, \ref{fig8}, \ref{fig9} and \ref{fig10}, the location of the shockwave as a function of time is obtained. An image cleaning algorithm programmed in MATLAB is used to remove the unwanted noise in the acquired images. A canny edge detection algorithm is used to identify the location of the shock front. An intensity scan is performed along the central axis of the driven tube. With the shock front location at different time instants, the velocity of the shock front is estimated. Figure \ref{fig11} shows the variation of the shock Mach number along the driven section of the three shock tubes when $P_{41}=15$ and driver gas is nitrogen. The prediction of the 1-D inviscid shock tube theory is also indicated as a dotted line in the graphs. The shock Mach number is computed using backward finite-difference of the shock front location at different time instants in the images. Since the region between the two threads (at a distance of 140mm and 160mm from diaphragm station) is common for Part 1 and Part 2, there are overlapping data points in this region. The shock Mach number increases to a peak value and then gradually decreases. The distance along the shock tube where the shock velocity reaches a peak value is called the shock formation distance and is represented by $x_f$ \citep{low1976}. The velocity and Mach number of the shockwave is represented by $V_S$ and $M_S$, respectively. The peak value of the velocity and Mach number is represented by $V_{Smax}$ and $M_{Smax}$. The variation of the shock Mach number is similar to that reported by \citet{glass1955}. The shockwave accelerates until the shock formation distance and then gradually loses strength. From the plots in figure \ref{fig11}, $M_{Smax}$ for the 10mm shock tube is 1.71 at a distance of 93mm from diaphragm location ($x/D=9.3$). In the 6mm shock tube, $M_{Smax}$ is 1.62 at a distance of 44mm ($x/D=7.3$) and for the 2mm shock tube, $M_{Smax}$ is 1.42 at 18mm ($x/D=9$). A strong conclusion that can be made is that to obtain the same Mach number with decreasing shock tube diameters, the diaphragm pressure ratio, $P_{41}$, must be increased. The time stamp, $t$, is represented as a dimensionless parameter, $t^*$, defined as,

\begin{equation}
	t^* = \frac{t.a_1}{D}
	\label{e2_3_t*}
\end{equation}

The figure \ref{fig11}d shows the plot between the dimensionless parameters, $t^*$ and $x/D$, for the three shock tubes. The variation of the dimensionless quantities plotted in figure \ref{fig11}d is similar for the three shock tubes. Since the exact time of the diaphragm rupture is not known in the Part2 visualizations, the arrival time of the shockwave cannot be correlated with the Part1 visualizations. The shock tube's driven section is divided into two regions based on the variation of the shock Mach number. The region associated with the shockwave acceleration is called the shock formation region, and the region beyond the shock formation distance is referred to as the shock propagation region. The experimental results are explained in the subsequent sections by closely investigating the flow phenomena due to the finite diaphragm rupture process and viscous effects due to the shock tube walls.

	\section{Shock formation region in the shock tube}\label{sec:shock_form}
Various flow models have been reported in the literature that incorporates the diaphragm opening process, the shock formation distance, and the diaphragm opening time. In this section, a model has been developed to validate the experimental findings.  The flow phenomena are explained as a result of the finite rupture time of the diaphragm.

	\subsection{Modeling the shock formation process}

 \begin{figure}
  \centerline{\includegraphics[scale=0.7]{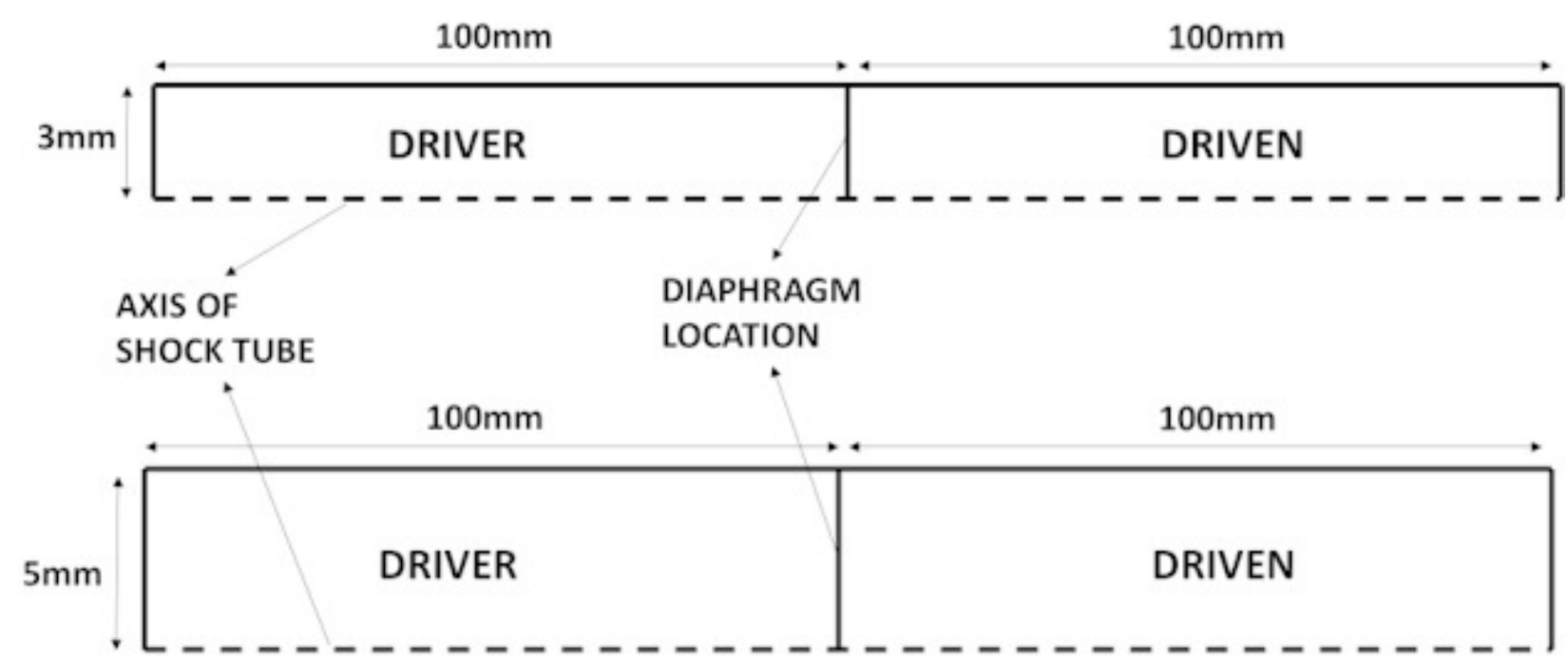}}
  \caption{A schematic diagragm of the computational domain for the 6mm shock tube (top) and the 10mm shock tube (bottom) used for the simulations.}
\label{fig12}
\end{figure}

\begin{figure}
  \centerline{\includegraphics[scale=0.7]{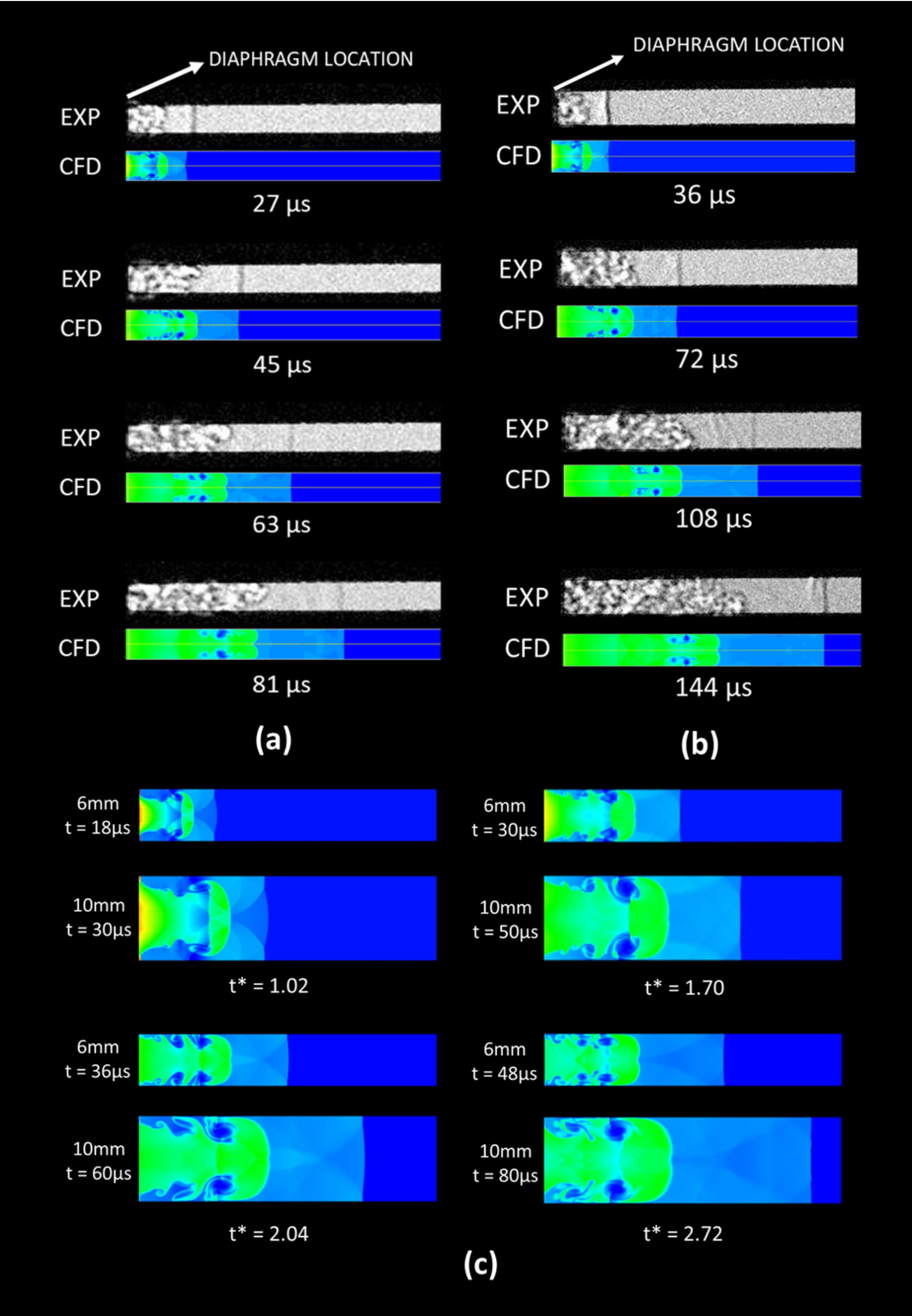}}
  \caption{Comparison of the experiments and simulations for the (a) 6mm shock tube and (b) 10mm shock tube. (c) Comparison between the CFD results for the 6mm and 10mm shock tubes. $P_{41} = 15$ and Nitrogen is driver gas.}
\label{fig13}
\end{figure}

As observed in the experiments, the shock formation region is dominated by the wave reflections and interactions due to the finite rupture time of the diaphragm. Therefore, a two-dimensional inviscid simulation is performed to model the region around the diaphragm location in the miniature shock tubes. Since the 2mm shock tube's shadowgraphs reveal very little in terms of the wave interactions behind the shock front, simulations are performed for the 6mm and 10mm shock tube. Two-dimensional axisymmetric models are used to simulate the 6mm and the 10mm shock tube numerically. The details of the models are shown in figure \ref{fig12}. An inviscid flow is considered. The simulations are run in a commercial solver, ANSYS FLUENT 13.0. The driver and driven gases are considered to obey ideal gas law. The flux component of the governing equation was discretized using the Roe-FDS scheme. A first-order implicit scheme was used for the temporal discretization. To simulate the diaphragm's gradual opening in the shock tube, the model proposed by \citet{Arun2013_1} is used. The diaphragm rupture process is assumed to follow a quadratic mathematical function \citep{Matsuo2007}. The following equation represents the quadratic mathematical function,

\begin{equation}
	t_o = \{ (r/R)^2T \}
	\label{e3_dia}
\end{equation}

where $r$ represents the opening radius at some arbitrary time $t_o$, $R$ represents the initial radius of the diaphragm, and $T$ represents the total diaphragm opening time. For the 6mm shock tube simulation, the diaphragm is divided into 30 parts of 0.1mm each. The diaphragm is divided into 50 parts of 0.1mm each in the 10mm shock tube. The total diaphragm opening time is divided into discrete time steps based on equation \ref{e3_dia}, and the diaphragm opening radius at each time step is found. Portions of the diaphragm are removed in the simulation to match the opening radius at any time instant. To estimate the total opening time of the diaphragm for the shock tubes, equation \ref{e1_op} is used. The thickness of the diaphragm is considered to be 40$\mu m$. The density of the cellophane diaphragm is taken as 1500 $kg/m^3$. The calculated values of the diaphragm opening time for the 10mm shock tube and 6mm shock tube is 37$\mu s$ and 28$\mu s$, respectively.

\begin{figure}
  \centerline{\includegraphics[scale=0.4]{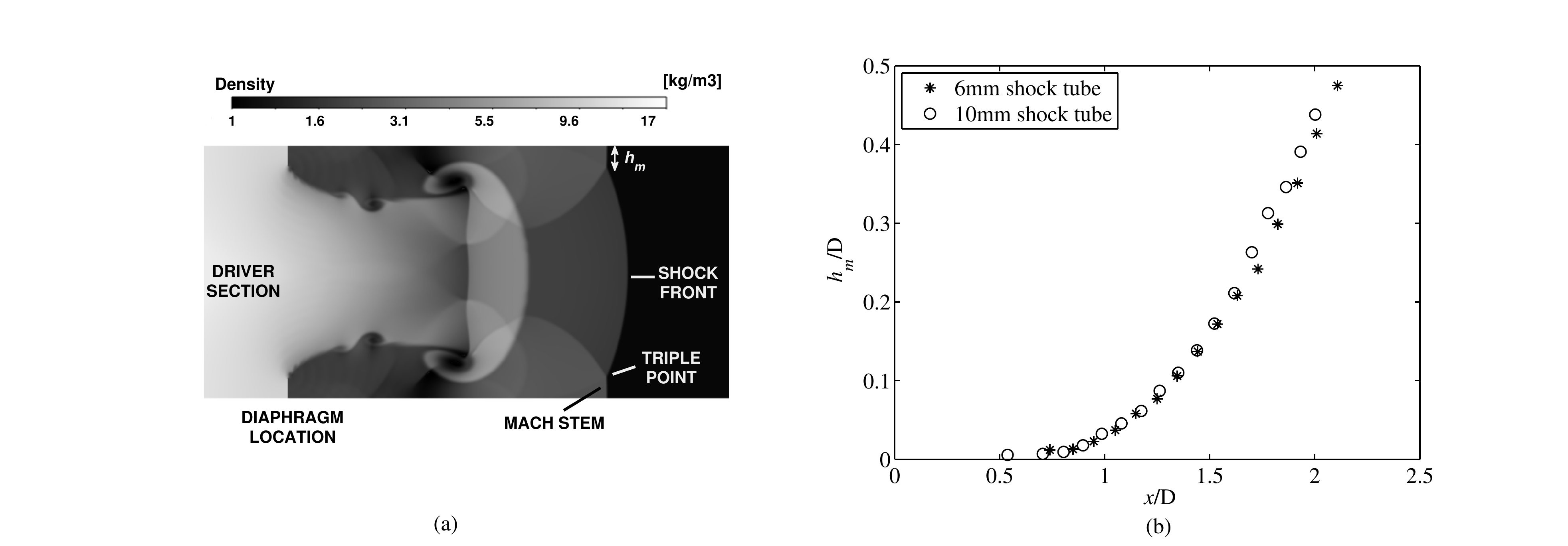}}
  \caption{(a) Density contour of the 10mm shock tube showing the Mach stem and triple point. The Mach stem height, $h_m$, is indicated in the figure. (b) Plot comparing the growth of the Mach stem in the 6mm and 10mm shock tube.}
\label{fig14}
\end{figure}

\begin{figure}
  \centerline{\includegraphics[scale=0.55]{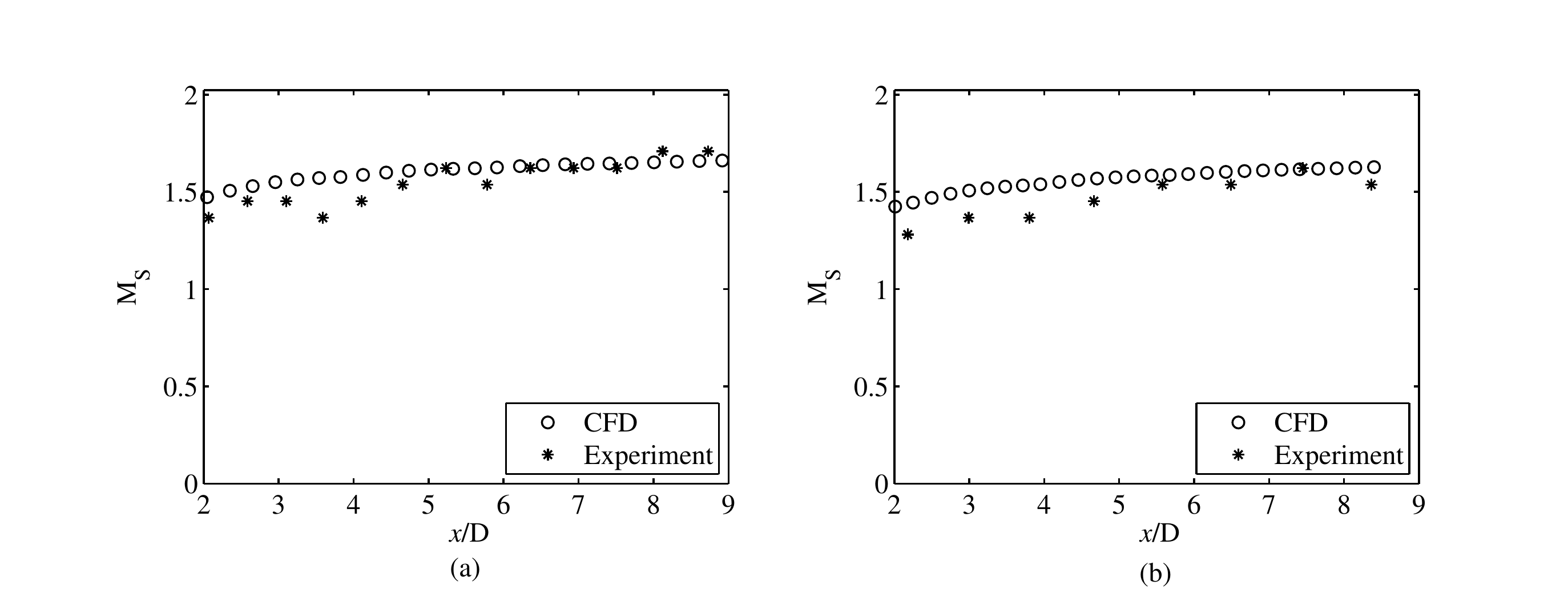}}
  \caption{Comparison between the shock Mach number obtained from experiment and CFD simulations for (a) the 10mm shock tube and (b) the 6mm shock tube.}
\label{fig15}
\end{figure}

Figures \ref{fig13}a and \ref{fig13}b show the comparison between the experimental shadowgraph and the density contours obtained using simulations for the 6mm and 10mm shock tube, respectively. From the observed contours, useful observations are made in the flow phenomena immediately after the diaphragm rupture is initiated. These figures closely explain the shock formation in practical scenarios where the shockwave forms at a finite distance from the diaphragm station due to the diaphragm's non-instantaneous rupture. The position of the shock front and contact surface is captured with reasonable accuracy in the computations. It is seen that the initial shape of the shock front is spherical. The reflection of the spherical shock front from the walls leads to the formation of a Mach stem. Vortices are formed at the contact surface region that leads to mixing between the driver and driven gases. There are oblique structures between the shock front and the contact surface immediately after the diaphragm ruptures, catching up with the initial shock front. The shock front becomes planar at 63$\mu s$ (corresponds to $t^*=3.6$) for the 6mm shock tube and at 108$\mu s$ (corresponds to $t^*=3.7$) for the 10mm shock tube. Therefore, the shockwave in the two shock tubes become planar at the same dimensionless time. The shape of the shock front observed in the 10mm shock tube case is captured well in the computations. The figure \ref{fig13}c shows the comparison of 6mm and 10mm shock tube simulations at the same $t^*$ after the simulation start. The wave phenomena behind the shock front are similar for both 6mm and 10mm shock tube cases at the same $t^*$ values. The similarity highlights that analogous flow features are observed at the same dimensionless time stamps. The figure \ref{fig14}a shows a snapshot of the density contour obtained in the simulation for the 10mm shock tube. The formation of the triple point and the Mach stem is evident in the figure. The height of the Mach stem ($h_m$) is also indicated. The trajectory of the triple point and the height of the Mach stem is measured in the simulations. The scaled Mach stem height, $h_m/D$, increases proportionally with the dimensionless parameter, $x/D$, for both the 6mm and 10mm shock tube cases (see figure \ref{fig14}b). Therefore, the shock formation process in the different diameters can be correlated based on the scaled parameters. The derivation of the correlation is elaborated in the following section. Figure \ref{fig15} shows the variation of the shock Mach number along the driven section of the shock tube in the experimental and computational results. The shockwave location is tracked in the computational results based on the pressure jump at the shock front. The shock Mach number is computed from the location of the shockwave at different time instants. The shock Mach number at the shock formation distance predicted by the numerical simulations are similar to those observed in the experiments. These numerical simulations show that the wave interactions dominate the shock formation process resulting from the shock tube walls' reflections.

\begin{figure}
  \centerline{\includegraphics[scale=0.35]{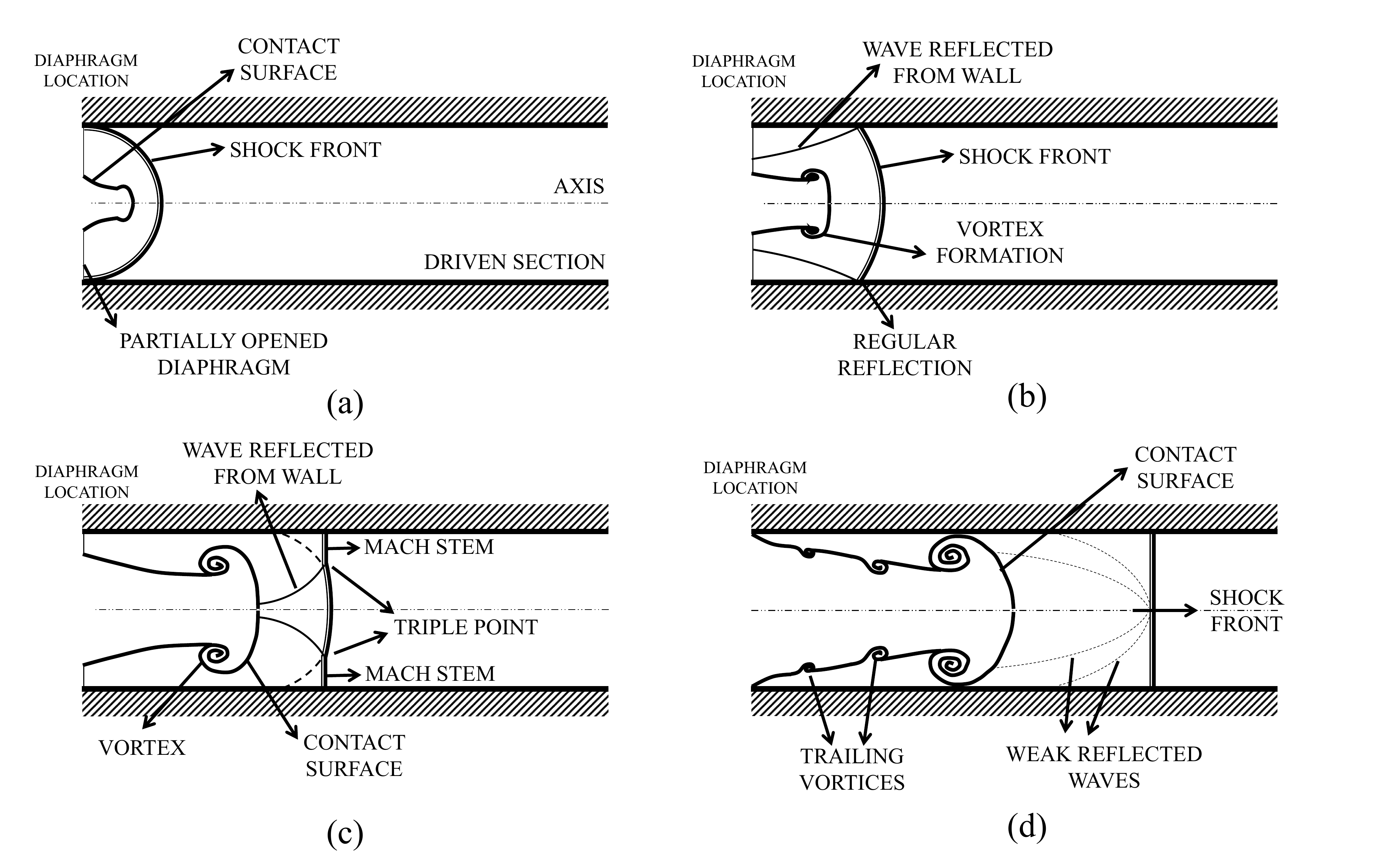}}
  \caption{Schematic diagrams showing the flow evolution and formation of the shockwave in the driven section as the diaphragm progressively opens from the center of the shock tube.}
\label{fig16}
\end{figure}

Figure \ref{fig16} shows the flow features very close to the diaphragm based on the experiments and computations' observations. As soon as the diaphragm opens/tears from the center, the high-pressure driver gas gushes into the low pressure-driven gas, preceded by a shock wave that expands spherically in the space bounded by the walls of the shock tube (see Figure \ref{fig16}a). The spherical nature of the shock front is attributed to a small portion of the diaphragm being opened. The spherical shock front expands in all directions inside the tube and reflects off the shock tube wall (see Figure \ref{fig16}b). The flow of high-pressure driver gas past the partially opened diaphragm into an initially stationary low-pressure gas leads to the formation of counter-rotating vortex rings. The reflected shock moves towards the axis and interacts with the curved contact surface, vortex rings, and the shocked gas region. The regular reflection of the reflected wave transforms into a Mach reflection as it catches up with the initial curved shock front, as shown in figure \ref{fig16}c. Such a reflection leads to the formation of a Mach stem and a triple point. This phenomenon is similar to flow features observed when spherical shockwave reflections occur on interaction with a flat surface \citep{ben2007shock}. A typical example is when explosions occur at a specific height from the ground, and the blast wave interacts with the surface \citep{needham2010}. The vortex rings formed at the contact surface become larger and move towards the shock tube wall. This phenomenon leads to mixing the driver and the driven gases as more low-pressure gas gets trapped behind the enlarging vortex. As the diaphragm opens up completely, trailing vortices are formed at the contact discontinuity (see Figure \ref{fig16}d). As time progresses, the Mach stem increases in height, the triple point moves closer to the center of the shock tube, and the shock front becomes planar. When the triple point from opposite sides of the flow meet, the reflected waves interact and form a wave system of weak compression waves that eventually catch up with the moving shock front. The shock front accelerates in the shock formation phase until the transverse wave reflections die down, and vortices at the contact discontinuity are shed. Quantitative velocity measurements and robust simulations planned in the future will help in validating these findings.

	\subsection{Correlation to predict shock Mach number in shock formation region}
The comparison between the experimental results and computations shows that the shock formation process is dominated by wave interactions due to the finite time for diaphragm rupture. The effect of the diaphragm's finite rupture time is that there are many reflected waves as a result of the initial compression wave from the diaphragm location that coalesces to form a single shock front. The first compression wave originating from the diaphragm location travels at the sound speed in the driven gas, i.e., $a_1$. The shock front keeps gaining speed as it propagates down the tube as compression waves catch up. Therefore, the velocity of the shock front very close to the diaphragm location is $a_1$. Therefore, the minimum value of the shock front in the shock formation region is $a_1$. The increment in the shockwave velocity is dependent on initial conditions of the gases present on either side of the diaphragm, physical length scales, and the mechanical properties of the diaphragm. The dependent quantities are:

\begin{itemize}
    \item Diameter of the shock tube ($D$)
    \item Pressure in the driver and driven section ($P_4$ and $P_1$)
    \item Speed of sound in the driver gas ($a_4$) and driven gas ($a_1$)
    \item Diaphragm opening time ($t_{op}$)
    \item Shock formation distance ($x_f$)
\end{itemize}

The diaphragm opening time is dependent on the material properties of the diaphragm and the initial pressure difference across the diameter. To understand the variation of the shockwave velocity with these parameters, experimental data reported by \citet{low1976,ikui1979,ikui1969} is considered along with the present experimental findings in the 2mm, 6mm, and 10mm shock tubes. Table \ref{t4_form} shows the consolidated data for all the experimental conditions. The parameter $a_{41}$ is the ratio of $a_4$ and $a_1$. The increase in shockwave velocity may be represented in terms of the dependent quantities in the following manner.

\begin{equation}
	V_{Smax}-a_1 = f(x_f,D,P_{41},a_1t_{op},a_{41})
	\label{e4_for1}
\end{equation}

	\begin{table}
		\begin{center}
			\def~{\hphantom{0}}
			\begin{tabular}{lccccccccccc}
				 & \multicolumn{5}{c}{Initial experimental conditions} &  & \multicolumn{3}{c}{Observed data} & \multicolumn{2}{c}{Correlation} \\
                Data source & Driver & Driven & $a_{41}$ & $P_{41}$ & D & $M_{Si}$ & $t_{op}$ & $x_f$ & $M_{Smax}$ & A & $M_{Smax}$ \\
                 & & & &  & $(mm)$ &  & $(\mu s)$ & $(mm)$ &  &  & \\

Janardhanraj & N$_2$  & Air & 1.009 & 15   & 10 & 1.73 & 37 & 93 & 1.71 & 0.21 & 1.72 \\
Janardhanraj & N$_2$  & Air & 1.009 & 15   & 6  & 1.73 & 28 & 44 & 1.62 & 0.21 & 1.62 \\
Janardhanraj & N$_2$  & Air & 1.009 & 15   & 2  & 1.73 & 16 & 18 & 1.42 & 0.16 & 1.42 \\
Shtemenko    & He  & Air & 2.910 & 45   & 46 & 3.01  & 420 & 500 & 2.08 & 0.12 & 2.06 \\
Shtemenko    & He  & Air & 2.910 & 45   & 46 & 3.01 & 680 & 800  & 1.97 & 0.15 & 2.39 \\
Shtemenko    & He  & Air & 2.910 & 45   & 46 & 3.01 & 750 & 900  & 1.73 & 0.08 & 1.75 \\
Rothkopf     & H$_2$  & Air & 3.671 & 9500 & 52 & 10.49 & 220 & 950  & 9.25 & 0.25 & 9.42 \\
Rothkopf     & H$_2$  & Air & 3.671 & 9500 & 52 & 10.49 & 375 & 1040 & 8.58 & 0.26 & 8.40 \\
Rothkopf     & He  & Air & 2.910 & 9500 & 52 & 6.95 & 220 & 800  & 7.11 & 0.25 & 7.12 \\
Rothkopf     & He  & Air & 2.910 & 9500 & 52 & 6.95 & 375 & 940  & 6.68 & 0.26 & 6.58 \\
Rothkopf     & Ar  & Air & 0.922 & 9500 & 52 & 2.86 & 220 & 800  & 3.47 & 0.32 & 3.48 \\
Rothkopf     & Ar  & Air & 0.922 & 9500 & 52 & 2.86 & 375 & 800  & 3.32 & 0.37 & 3.32 \\
Ikui         & Air & Air & 1.000 & 10   & 38 & 1.61 & 900 & 600  & 1.55 & 0.25 & 1.54 \\
Ikui         & Air & Air & 1.000 & 34   & 38 & 2.01 & 900 & 1000 & 1.60 & 0.19 & 1.60 \\
Ikui         & Air & Air & 1.000 & 100  & 38 & 2.37 & 900 & 1000 & 1.80 & 0.23 & 1.81 \\
			\end{tabular}
			\caption{Comparison between values predicted by correlation and experimentally obtained values in shock formation region.}
			\label{t4_form}
		\end{center}
	\end{table}

The term $a_1t_{op}$ represents the distance traveled by a compression wave with a velocity equal to the speed of sound in the driven gas. This characteristic distance helps represent the diaphragm opening time in terms of a length scale. The experimental results of \citet{low1976} helps in finding the variation of the quantities $a_1t_{op}$ and $a_{41}$ with $V_{Smax}$ while other parameters are kept constant. Similarly, \citet{ikui1979} gives the variation of $P_{41}$ with $V_{Smax}$ when the other parameters are kept a constant. Once these relationships are determined, the variation of $x_f$ with $V_{Smax}$ can also be found. It is seen that $V_{Smax}$ is directly proportional to $x_f$ $P_{41}$ $a_{41}$ but inversely proportional to $a_1t_{op}$. As a result of this analysis and combining the terms, the relationship between the quantities can be written as,

\begin{equation}
	\frac{V_{Smax}-a_1}{a_1} = f\left(\frac{x_f}{D},P_{41},\frac{D}{a_1t_{op}},a_{41}\right)
	\label{e5_for2}
\end{equation}

From figure \ref{fig11}, the variation of the shockwave velocity with distance is observed to follow a parabolic trend in the shock formation region. The quantities in the previous equation maybe considered to follow a power relationship and the relation can be written as,

\begin{equation}
	\frac{V_{Smax}-a_1}{a_1} = A. \left(\frac{x_f}{D}\right)^{C1}.(P_{41})^{C2}.\left(\frac{D}{a_1t_{op}}\right)^{C3}.(a_{41})^{C4}
	\label{e6_for3}
\end{equation}

where $A$ is constant of proportionality and $C1$, $C2$, $C3$, $C4$ are power constants. The values of the power constants are determined by substituting the values of the quantities given in the table \ref{t4_form}. After determining the values of the constants the equation \ref{e6_for3} is given by,

\begin{equation}
	M_{Smax}-1 = A. \left(\frac{x_f}{D}\right)^{0.5}.(P_{41})^{0.1}.\left(\frac{D}{a_1t_{op}}\right)^{0.4}.a_{41}
	\label{e7_for4}
\end{equation}

where the value of the constant $A$ lies in the range 0.08 - 0.37. The performance of equation \ref{e7_for4} with previously reported results is shown in table \ref{t4_form}. The scatter in the value of constant $A$ is reasonable, considering the wide range of supplementary variables shown in the table. The performance of this correlation with experimentally observed shock Mach number at different locations in the driven section of the shock tubes is reported in section \ref{sec:dis}.

	\section{Shock propagation region in the shock tube}\label{sec:shock_prop}
	\subsection{Attenuation due to wall effects}

\begin{figure}
  \centerline{\includegraphics[scale=0.6]{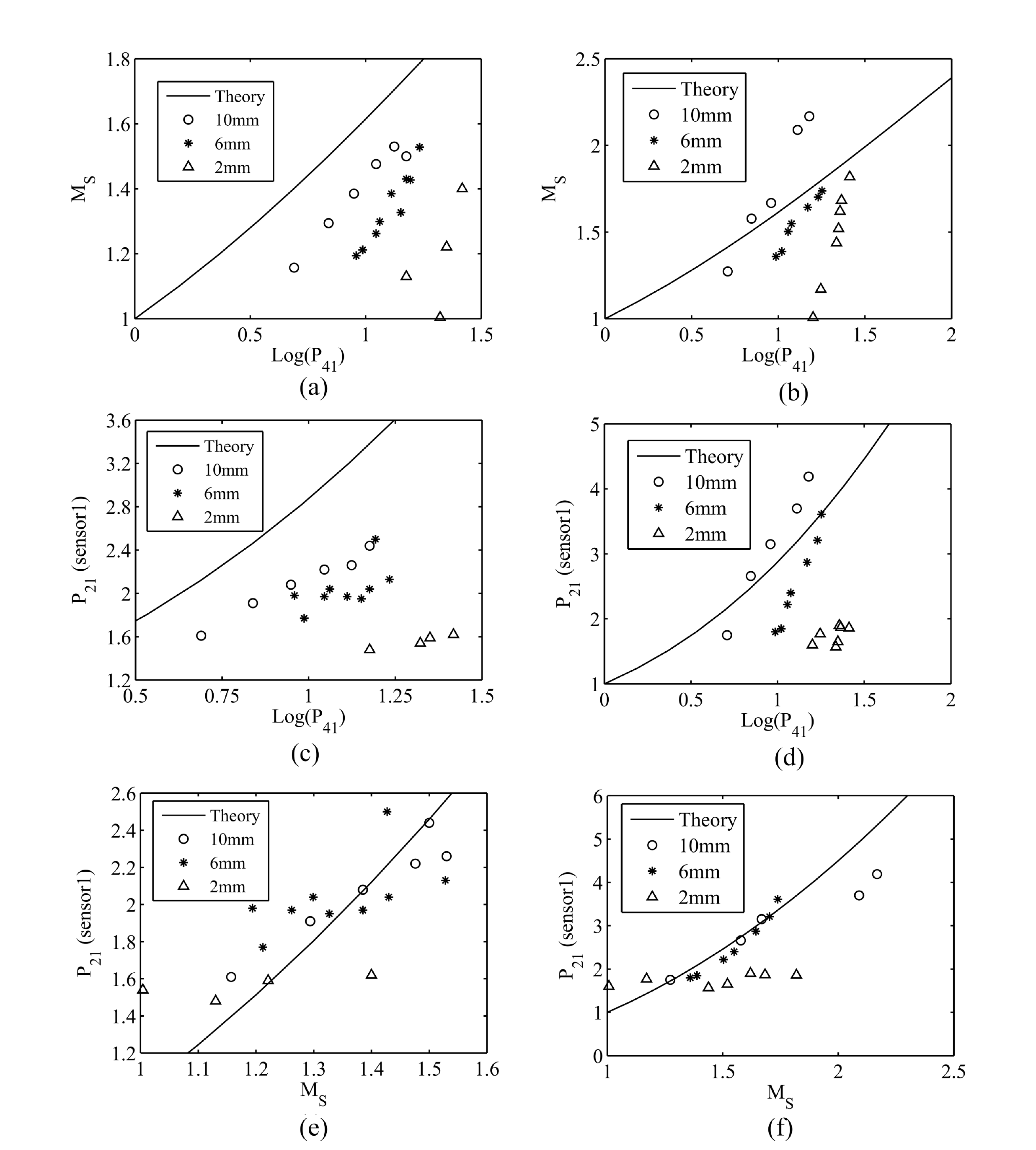}}
  \caption{Variation of the measured shockwave parameters. (a), (c), and (e) are results for nitrogen driver. (b), (d), and (f) are results for helium driver. The predictions of the 1-D inviscid shock tube theory is also indicated as a solid line in the graphs.}
\label{fig17}
\end{figure}

The one-dimensional inviscid shock tube theory is routinely used for calculating the various output shockwave parameters when underlying assumptions are valid, for example, in large shock tubes. However, this theory does not predict the fluid properties accurately in miniature shock tubes. Nonetheless, the values of the shockwave parameters obtained from experiments in the present work are compared with those obtained from the one-dimensional inviscid shock tube theory to observe the trends. Figure \ref{fig17} shows the experimental results plotted against the prediction of the one-dimensional inviscid shock tube theory for all the three shock tubes and driver gas configurations. The data points for the 2mm shock tube are farther away from the ideal shock relations as compared to the 6mm and the 10mm case (see Figure \ref{fig17}a). Therefore, as the internal cross-section is reduced from 10mm to 2mm, a higher diaphragm pressure ratio is required to achieve the same particular shock Mach number. From figure \ref{fig17}c, it is also evident that as the internal cross-section is reduced, a higher diaphragm pressure ratio is required to achieve the same pressure behind the shockwave. The figure \ref{fig17}e shows that in the 6mm and 10mm shock tube, the ratio $P_{21}$ increases with the shock Mach number. Similar observations are made when helium is used as driver (see Figures \ref{fig17}b, \ref{fig17}d and \ref{fig17}f). The relation between $P_{21}$ and $M_S$ matches the predictions of ideal theory (except for higher $M_S$ in the 10mm case). These observations show that the experimental findings match the expected trends in miniature shock tubes.

\begin{figure}
  \centerline{\includegraphics[scale=0.7]{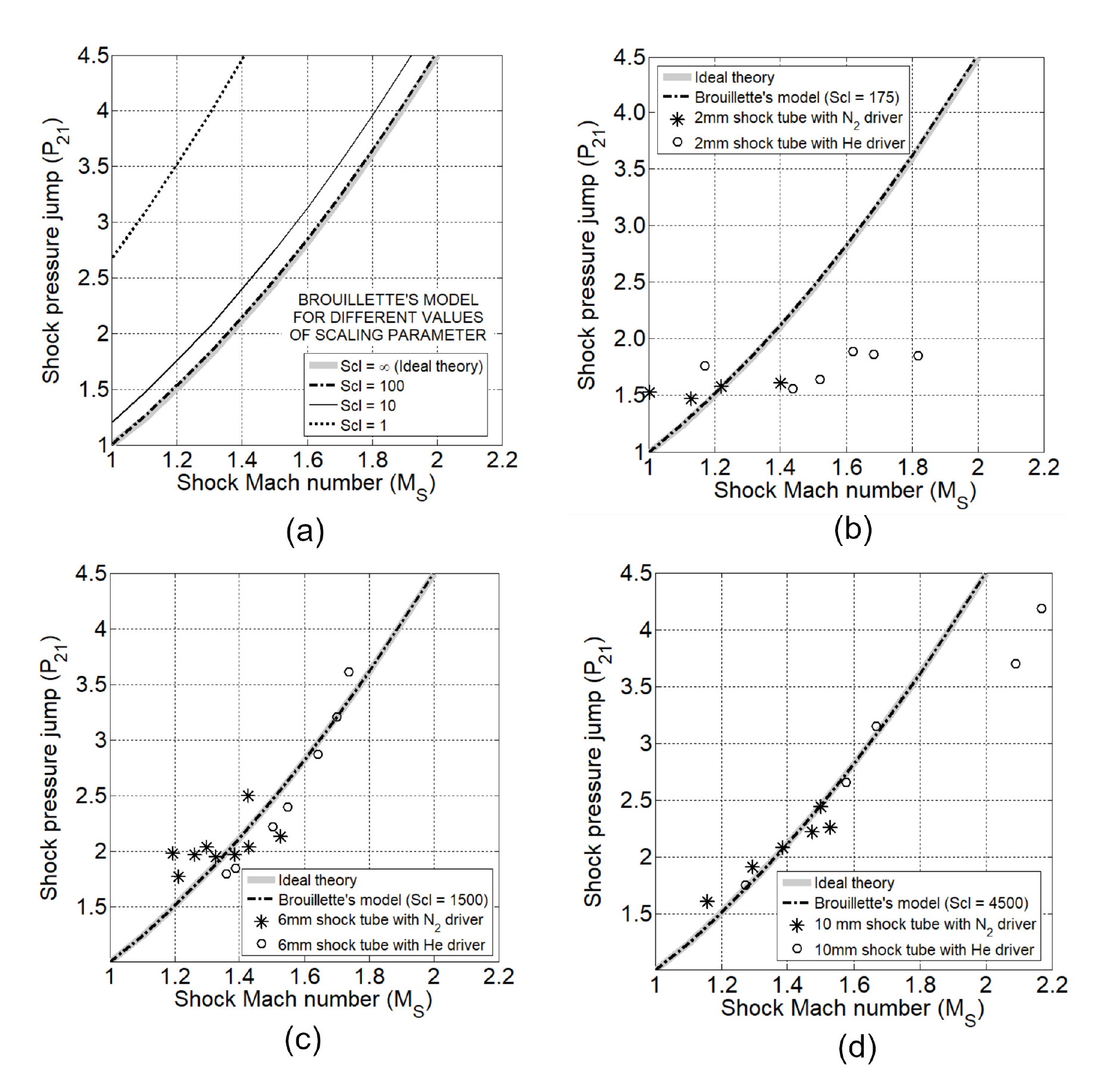}}
  \caption{Plots showing the relation between $P_{21}$ and $M_S$ (a) Comparison of Brouillette's model for different values of scaling parameter. (b) Results for the 2mm shock tube (c) Results for the 6mm shock tube (d) Results for the 10mm shock tube.}
\label{fig18}
\end{figure}

The attenuation of shockwaves due to boundary layer development is well-known and researched for many decades. The development of a scaling parameter has been explored numerically and experimentally for shockwave propagation through microchannels \citep{Broui2003}. This model accounts for the effect of length scales through molecular diffusion phenomena by parameterizing the shear stress and heat flux at the shock tube wall. As proposed by \citet{Broui2003}, the scaling parameter is defined as $Re'.D/4L$ where $Re'$ is the characteristic Reynolds number, and $ L $ is the characteristic length. The characteristic length is defined as the distance between the shockwave and the contact surface. The 1-D shock relations are used to find the distance between the contact surface and the shockwave for the present experiments' initial conditions. The scaling parameter values lie in the range of $175 < Scl < 5879$ for the present experimental conditions. Figure \ref{fig18}a compares the predictions for different values of the scaling parameter. It is clear that for $Scl > 100$, the model is the same as the one-dimensional inviscid shock tube theory. Figures \ref{fig18}b, \ref{fig18}c and \ref{fig18}d show the experimentally obtained variation of $P_{21}$ with $M_S$ as compared to the predictions using Brouillette's model for the 2mm, 6mm and 10mm shock tube respectively. In reality, the distance between the contact surface and the shock front is less than the value given by one-dimensional shock relations. Therefore, the scaling parameter is higher than the values taken for the present analysis. It is observed that Brouillette's model is closer to the ideal theory for large values of the scaling parameter. Therefore, the shockwave attenuation cannot be predicted by the model proposed by \citet{Broui2003} when the scaling parameter is greater than 100, and the shock formation region is present in the shock tube.

Using the scaling parameter, \citet{Zeitoun2015} presented a power law correlation to predict of attenuation in the shock Mach number in laminar and turbulent flows in the shock tube. The relation is given by,

\begin{equation}
  -\left(\frac{M_S-M_S^i}{M_S^i}\right)= C_a(Scx)^B
  \label{e8_att1}
\end{equation}

where $M_S^i$ corresponds to the initial shock Mach number, $C_a$ is the attenuation parameter, $Scx$ is the local scaling ratio, and $B$ is -1/7 for turbulent flow regime. The local scaling ratio is a function of the local position of the shockwave $x$ given by $Scx=(Re'D)/4x$. It was reported that the attenuation parameter gradually increases from zero to a value of 0.6, where maximum attenuation occurs. The maximum attenuation occurs at a distance that corresponds to 200 diameters. For the present scenario, it is observed that the shockwave velocity increases until the shock formation distance and subsequently decreases. Therefore, if the shock propagation region is defined as the region where the wall effects dominate the attenuation, then the shockwave's initial velocity can be taken equal to the maximum shock velocity reached during the end of the shock formation region. The correlation in equation \ref{e8_att1} can be modified appropriately to support this assumption. The derivation and performance of the modified correlation are presented in the subsequent section.

	\subsection{Correlation to predict shock Mach number in propagation region}
 As mentioned in the previous section, the maximum Mach number attained by the shockwave at the end of the shock formation region is considered as the starting point for the correlation that is developed for the shock propagation region. Therefore, equation \ref{e8_att1} is modified as follows,

 \begin{equation}
    M_S=M_{Smax}(1-C_a.(Scx)^B)
    \label{e9_att2}
 \end{equation}

where $M_S$ represents the Mach number of the shockwave in the propagation region, $M_Smax$ is the peak Mach number reached by the shockwave at the end of the shock formation region, $C_a$ is the attenuation parameter and $B$ is -1/7. $Scx$ in equation \ref{e9_att2} is the scaling parameter at the local position of the shockwave represented by $Scx=(Re'D)/(4(x-x_f))$ where $x_f$ is the shock formation distance. The local scaling parameter is changed because the shockwave's effective distance in the propagation region is $(x - x_f)$. The correlation between the velocity and the shockwave's local position, as represented in equation \ref{e9_att2}, is used to compare with the experimental data. The experimental data presented by Shtemenko and reported in \citet{ikui1969} is also used for comparison. Table \ref{t5_att} shows the values of the Mach number obtained using the experimentally measured pressure signals and compared them with those computed using the correlation. The value of $x$ that corresponds to the first pressure transducers' location in the driven section is used. It is observed that the attenuation parameter takes values in the range of 0.30 - 0.37 for the experimental data. It was reported that the value of the attenuation parameter increases from zero to a value of 0.6 within about 300 times the tube diameter for a turbulent flow regime. It can be seen in the table that $(x - x_f)/D$ lies within 300. Therefore, the values of 0.30 - 0.37 are reasonable. Another important point is that the value of the attenuation parameter increases with the scaled distance. Therefore, the correlation between shock Mach number and local position of the shockwave represented in equation \ref{e9_att2} works well for the propagation region.

	\begin{table}
		\begin{center}
			\def~{\hphantom{0}}
			\begin{tabular}{lccccccccccc}
				Data source	& $D$ & $P_{41}$ & $x_f$ & $M_{Smax}$ & $Re’$ & $x$ & $\frac{x-x_f}{D}$ & $Scx$ & $C_a$ & $M_S$(corr) &	 $M_S$(e)\\[3pt]
                 & $(mm)$ &  & $(mm)$ &  &  & $(mm)$ &  &  &  &  &  \\
Janardhanraj & 10 & 15 & 93  & 1.71 & 225741 & 311  & 23    & 2454 & 0.36 & 1.51 & 1.50 \\
Janardhanraj & 6  & 15 & 44  & 1.62 & 135444 & 311  & 44.5  & 761  & 0.36 & 1.40 & 1.39 \\
Janardhanraj & 2  & 15 & 18  & 1.42 & 45148  & 311  & 145.5 & 78   & 0.37 & 1.14 & 1.13 \\
Shtemenko    & 46 & 45 & 500 & 2.08 & 68919  & 1000 & 10.9  & 1585 & 0.34 & 1.84 & 1.82 \\
Shtemenko    & 46 & 45 & 800 & 1.97 & 68919  & 1000 & 4.3   & 3963 & 0.33 & 1.77 & 1.76 \\
Shtemenko    & 46 & 45 & 900 & 1.73 & 68919  & 1000 & 2.2   & 7962 & 0.30 & 1.59 & 1.59 \\
			\end{tabular}
			\caption{Comparison between values predicted by correlation and shock Mach number obtained experimentally using pressure transducers.}
			\label{t5_att}
		\end{center}
	\end{table}

    \section{Discussions}\label{sec:dis}

The relations represented in equations \ref{e7_for4} and \ref{e9_att2} are used to predict the shock Mach number at different locations in the driven section plotted in figure \ref{fig11}. Equation \ref{e7_for4} can be represented in terms of the local position of the shockwave $x$ as follows,

\begin{equation}
	M_{S}-1 = A. \left(\frac{x}{D}\right)^{0.5}.(P_{41})^{0.1}.\left(\frac{D}{a_1t_{op}}\right)^{0.4}.a_{41}
	\label{e10_com1}
\end{equation}

\begin{figure}
  \centerline{\includegraphics[scale=0.6]{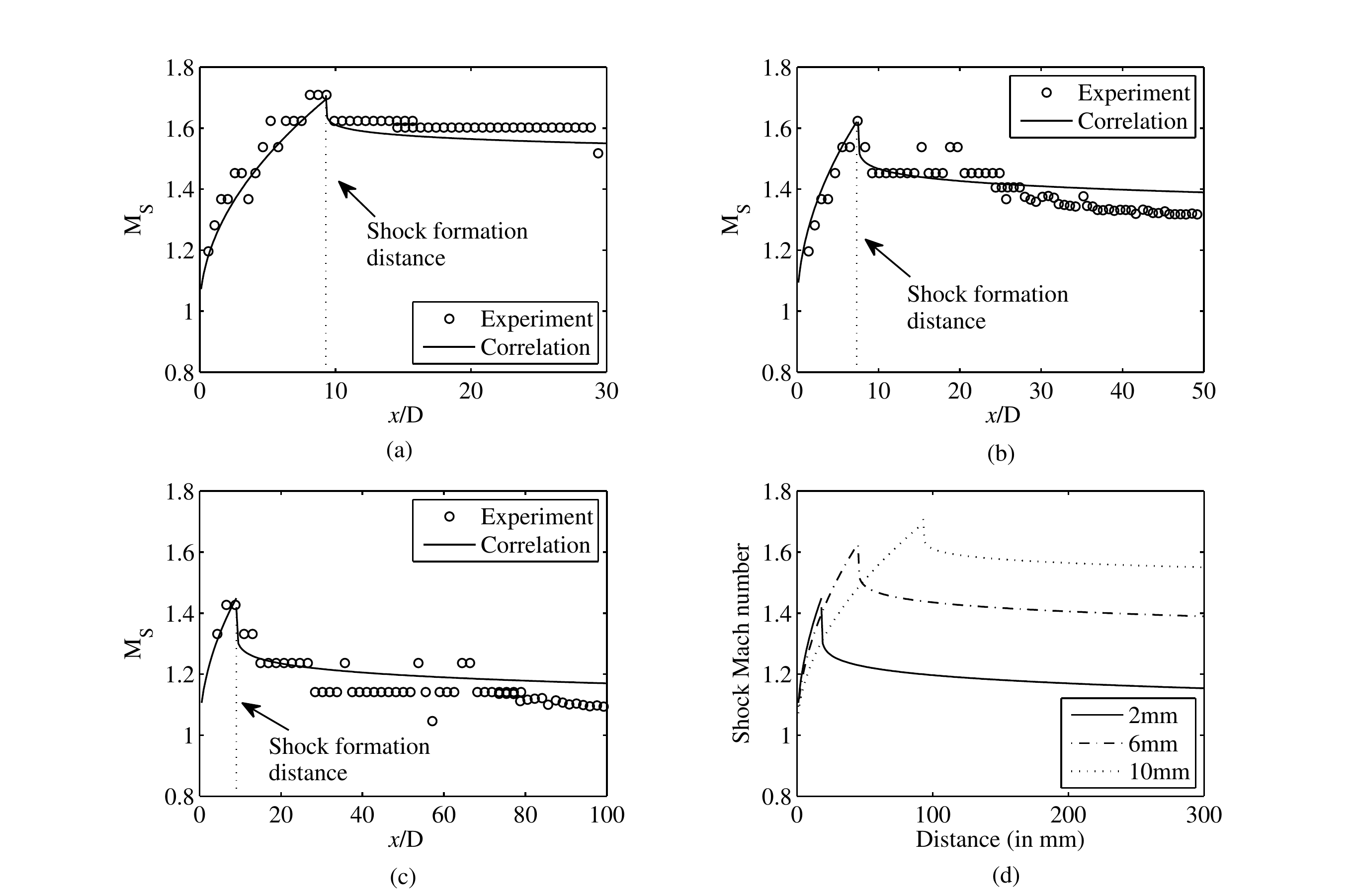}}
  \caption{Plots showing the comparison of the correlation with the experimental data points in the driven section of the (a) 10mm shock tube (b) 6mm shock tube and (c) 2mm shock tube. (d) A plot showing the distance-time graph of the shockwave trajectory obtained from correlations in the 10mm, 6mm and, 2mm shock tubes. $P_{41}=15$ and nitrogen is used as driver gas.}
\label{fig19}
\end{figure}

Figure \ref{fig19}a, \ref{fig19}b, and \ref{fig19}c show the total attenuation in shock Mach number as predicted by the correlations developed for the shock formation and the shock propagation regions for the 10mm, 6mm, and 2mm shock tubes. The correlation curves are obtained using the particular value of $C_a$ for each shock tube (shown in table \ref{t5_att}) and substituting the value of $x$ along the length of the driven tube. The shock formation distance is also indicated in the plots. The correlations predict the trend in the shock Mach numbers' variation for all the shock tubes very well. The figure \ref{fig19}d shows the comparison of the variation of the shock Mach number along the driven section for the three shock tubes. An important observation is that the acceleration of the shock front is highest in the 2mm shock tube and lowest in the case of the 10mm shock tube during the shock formation process. The results are consistent with those obtained in the simulations in section \ref{sec:shock_form}, which showed that the shock front travels faster in the 6mm shock tube as compared to the 10mm shock tube. Since the shock formation distance is small in the 2mm shock tube as compared to the other two hydraulic diameters, the peak Mach number of the shockwave is less for the 2mm shock tube than the other shock tubes. This observation shows the decrement in the shock Mach number in the shock formation region compared to the value predicted by the one-dimensional inviscid shock relations. The value of the shock Mach number predicted by the one-dimensional inviscid theory is 1.73. There is a steep drop in the velocity initially in the shock propagation region, as a gradual change in the attenuation parameter is not considered while plotting the correlation. But overall, the prediction of the shock Mach number in experiments using the developed correlations is satisfactory. The predictions using the correlations are mainly limited because the shock formation distance is not known a priori, but it is determined experimentally. A theoretical or empirical relationship to determine the shock formation distance accurately improves the predictions using the correlations. Thus, the flow in the miniature shock tubes can be divided into two regions: the shock formation region and the shock propagation region. The shock formation region is dominated by the wave reflections from the walls of the shock tube. The viscous effects are minimal in this region. The main parameters influencing the flow in this region are the hydraulic diameter of the shock tube ($D$), diaphragm pressure ratio ($P_{41}$), the speed of sound in the driver and driven gas ($a_4$ and $a_1$), the diaphragm opening time ($t_{op}$) and the shock formation distance ($x_f)$. The shock propagation region is dominated by the viscous effects and the boundary layer's development behind the shock front.

    \section{Conclusions}\label{sec:con}
A new table-top miniature shock tube system has been developed to understand the shock tube flow in 2mm, 6mm, and 10mm square cross-section shock tubes. This study gives more in-depth insights into the shockwave attenuation due to the shock formation and the shock propagation processes. The shock tubes are run at pressure ratios in the range 5-25 and driven section at initial ambient conditions so that the operating conditions are similar to those used in shockwave-assisted applications. Nitrogen and helium are used as driver gases to calibrate the shock tubes. The results from the experiments are compared with various numerical, empirical, and analytical models. The best agreement is obtained for an improved model suggested by Glass and Martin where the shockwave attenuation occurs in the two independent regions in the shock tube (1) The formation of the shockwave is dominated by waves generated due to the finite rupture time of the diaphragm and their reflections from the walls of the shock tube. The wave interactions happen in the smaller diameter shock tube earlier than in larger diameter shock tubes. The scaled Mach stem height increases proportionally with the scaled distance along the shock tube length for different diameter shock tubes. (2) After reaching the peak Mach number during the formation process, the propagation of the shockwave undergoes attenuation due to the formation of a turbulent boundary layer. The experimental findings indicate that the wave interactions and shock formation occur at the same dimensionless time in the shock tubes. Also, the maximum shock Mach number, which is reached at the shock formation distance, is higher for the 10mm shock tube case than the 2mm shock tube. New correlations have been developed to predict the shock Mach number in the shock formation and shock propagation region. Future experiments using the PIV (Particle Image Velocimetry) technique are planned to give valuable quantitative data close to the diaphragm station in the driver and driven section. Also, LES (Large Eddy Simulation) and DNS (Direct Numerical Simulation) studies of the shock tube flow due to the diaphragm's finite opening time will help validate the observations and experiments.\\

\noindent{\bf Acknowledgements\bf{.}} The authors would like to acknowledge the research grants from Defense Research and Development Organization (DRDO), India, towards this study. The authors are also thankful to the members of the Laboratory for Hypersonic and Shockwave Research (LHSR), Department of Aerospace Engineering, IISc, for their support and help during this study. \\

\noindent{\bf  Author ORCID\bf{.}} Janardhanraj S., https://orcid.org/0000-0002-1069-1306; Jagadeesh G., https://orcid.org/0000-0002-5495-9351\\

\appendix

\section{}\label{appA}

\setcounter{table}{0} \renewcommand{\thetable}{A.\arabic{table}}
Table \ref{t2_n2} shows the experimental results for different initial driver pressures for the 2mm, 6mm and 10mm square cross-section shock tubes when the driver gas is nitrogen. Table \ref{t3_he} shows the experimental results for different initial driver pressures for the 2mm, 6mm and 10mm square cross-section shock tubes when the driver gas is helium.

\begin{table}
		\begin{center}
			\def~{\hphantom{0}}
			\begin{tabular}{ccccccccccc}
                 \multicolumn{5}{c}{Experimental data} & & \multicolumn{2}{c}{1-D relations} & & \multicolumn{2}{c}{R-H relations} \\
				$P_{41}$ & D(mm) & $P_{21}$(sen1) & $P_{21}$(sen2) & $M_S(e)$ & \hspace{10pt} & $P_{21}(i)$ & $M_{S}(i)$ & \hspace{10pt} & $M_{S1}$ & $M_{S2}$ \\
4.9  & 10 & 1.61 & 1.54 & 1.16 &  & 2.10 & 1.39 &  & 1.23 & 1.21 \\
6.9  & 10 & 1.91 & 1.77 & 1.29 &  & 2.44 & 1.49 &  & 1.33 & 1.29 \\
8.9  & 10 & 2.08 & 1.91 & 1.39 &  & 2.71 & 1.57 &  & 1.39 & 1.33 \\
11.1 & 10 & 2.22 & 2.01 & 1.48 &  & 2.97 & 1.64 &  & 1.43 & 1.37 \\
13.3 & 10 & 2.26 & 2.09 & 1.53 &  & 3.19 & 1.69 &  & 1.44 & 1.39 \\
15.0 & 10 & 2.44 & 2.17 & 1.50 &  & 3.34 & 1.73 &  & 1.49 & 1.42 \\
9.1  & 6  & 1.98 & 1.75 & 1.19 &  & 2.74 & 1.57 &  & 1.36 & 1.28 \\
9.7  & 6  & 1.77 & 1.67 & 1.21 &  & 2.81 & 1.59 &  & 1.29 & 1.25 \\
11.1 & 6  & 1.97 & 1.74 & 1.26 &  & 2.97 & 1.63 &  & 1.35 & 1.28 \\
11.5 & 6  & 2.04 & 1.87 & 1.30 &  & 3.01 & 1.65 &  & 1.38 & 1.32 \\
12.9 & 6  & 1.97 & 1.74 & 1.39 &  & 3.15 & 1.68 &  & 1.35 & 1.28 \\
14.2 & 6  & 1.95 & 1.94 & 1.33 &  & 3.27 & 1.71 &  & 1.35 & 1.34 \\
15.0 & 6  & 2.04 & 1.97 & 1.39 &  & 3.34 & 1.73 &  & 1.38 & 1.35 \\
15.6 & 6  & 2.50 & 2.08 & 1.43 &  & 3.39 & 1.74 &  & 1.51 & 1.39 \\
17.1 & 6  & 2.13 & 1.99 & 1.53 &  & 3.51 & 1.77 &  & 1.40 & 1.36 \\
15.0 & 2  & 1.48 & 1.36 & 1.13 &  & 3.34 & 1.73 &  & 1.19 & 1.14 \\
21.0 & 2  & 1.54 & 1.45 & 1.00 &  & 3.79 & 1.84 &  & 1.21 & 1.18 \\
22.4 & 2  & 1.59 & 1.50 & 1.22 &  & 3.88 & 1.86 &  & 1.23 & 1.20 \\
26.2 & 2  & 1.62 & 1.53 & 1.40 &  & 4.11 & 1.91 &  & 1.24 & 1.21 \\
			\end{tabular}
			\caption{Experimental results for different pressure ratios ($P_{41}$) and nitrogen driver.}
			\label{t2_n2}
		\end{center}
	\end{table}

\begin{table}
		\begin{center}
			\def~{\hphantom{0}}
			\begin{tabular}{ccccccccccc}
                 \multicolumn{5}{c}{Experimental data} & & \multicolumn{2}{c}{1-D relations} & & \multicolumn{2}{c}{R-H relations} \\
				$P_{41}$ & D(mm) & $P_{21}$(sen1) & $P_{21}$(sen2) & $M_S(e)$ & \hspace{10pt} & $P_{21}(i)$ & $M_{S}(i)$ & \hspace{10pt} & $M_{S1}$ & $M_{S2}$ \\
                5.1  & 10 & 1.75 & 1.57 & 1.27 &  & 3.37 & 1.74 &  & 1.26 & 1.21 \\
7.0  & 10 & 2.66 & 2.35 & 1.58 &  & 4.21 & 1.93 &  & 1.53 & 1.44 \\
9.1  & 10 & 3.15 & 2.68 & 1.67 &  & 5.04 & 2.11 &  & 1.65 & 1.53 \\
12.9 & 10 & 3.70 & 3.11 & 2.09 &  & 6.36 & 2.36 &  & 1.78 & 1.64 \\
15.1 & 10 & 4.19 & 3.15 & 2.17 &  & 7.04 & 2.48 &  & 1.88 & 1.65 \\
9.7  & 6  & 1.80 & 1.60 & 1.36 &  & 5.26 & 2.15 &  & 1.28 & 1.22 \\
10.5 & 6  & 1.85 & 1.79 & 1.39 &  & 5.55 & 2.21 &  & 1.30 & 1.28 \\
11.4 & 6  & 2.22 & 2.08 & 1.50 &  & 5.86 & 2.27 &  & 1.41 & 1.37 \\
11.9 & 6  & 2.40 & 2.26 & 1.55 &  & 6.03 & 2.30 &  & 1.46 & 1.42 \\
14.8 & 6  & 2.87 & 2.50 & 1.64 &  & 6.95 & 2.47 &  & 1.58 & 1.48 \\
17.0 & 6  & 3.21 & 2.78 & 1.70 &  & 7.59 & 2.57 &  & 1.66 & 1.56 \\
17.9 & 6  & 3.61 & 3.10 & 1.74 &  & 7.84 & 2.62 &  & 1.76 & 1.64 \\
15.9 & 2  & 1.60 & 1.50 & 1.01 &  & 7.28 & 2.52 &  & 1.22 & 1.18 \\
17.6 & 2  & 1.77 & 1.61 & 1.17 &  & 7.76 & 2.60 &  & 1.27 & 1.22 \\
21.7 & 2  & 1.57 & 1.52 & 1.44 &  & 8.84 & 2.77 &  & 1.21 & 1.19 \\
22.3 & 2  & 1.65 & 1.62 & 1.52 &  & 8.99 & 2.80 &  & 1.23 & 1.22 \\
22.7 & 2  & 1.90 & 1.70 & 1.62 &  & 9.09 & 2.81 &  & 1.31 & 1.25 \\
23.2 & 2  & 1.87 & 1.73 & 1.68 &  & 9.21 & 2.83 &  & 1.30 & 1.26 \\
25.8 & 2  & 1.86 & 1.72 & 1.82 &  & 9.83 & 2.92 &  & 1.30 & 1.26 \\
			\end{tabular}
			\caption{Experimental results for different pressure ratios ($P_{41}$) and helium driver.}
			\label{t3_he}
		\end{center}
	\end{table}

\bibliographystyle{jfm}
\bibliography{shock}

\end{document}